# Investigating the causal effects of multiple treatments using longitudinal data: a simulation study

Emily Granger[1], Gwyneth Davies[2], Ruth H. Keogh[1]

[1]Department of Medical Statistics,
Faculty of Epidemiology and Population Health,
London School of Hygiene and Tropical Medicine,
Keppel Street, London, WC1E 7HT

[2] Population, Policy and Practice Research and Teaching Department,
UCL Great Ormond Street Institute of Child Health (UCL GOS ICH),
London WC1N 1EH, United Kingdom

**Abstract**

Many clinical questions involve estimating the effects of multiple treatments using observational data. When using longitudinal data, the interest is often in the effect of treatment strategies that involve sustaining treatment over time. This requires causal inference methods appropriate for handling multiple treatments and time-dependent confounding. Robins' Generalised methods (g-methods) are a family of methods which can deal with time-dependent confounding and some of these have been extended to situations with multiple treatments, although there are currently no studies comparing different methods in this setting. We show how five g-methods (inverse-probability-of-treatment weighted estimation of marginal structural models, g-formula, g-estimation, censoring and weighting, and a sequential trials approach) can be extended to situations with multiple treatments, compare their performances in a simulation study, and demonstrate their application with an example using data from the UK CF Registry.

## 1. Introduction

This study compares different analysis approaches for estimating the causal effects of multiple treatments using longitudinal observational data. There are many examples in the applied literature where the research question involves multiple treatments (or interventions, or exposures) [1-13]. Specific examples include: the effects of antiretroviral drug combinations on risk of cardiovascular disease [14]; the effects of intervening on depression, sleep duration, and leisure activity on cognitive impairment [15]; the joint effects of obesity and smoking on all-cause mortality [16]. Challenges may arise when estimating the effects of multiple treatments if, for example, at least one treatment has low prevalence, or the strength or cause of confounding differs for different treatment-outcome associations. In longitudinal data, issues could arise if few individuals remain on the same treatment over a long-term period (i.e., it is common to switch treatments).

In the simpler setting with one binary treatment, many statistical techniques for estimating causal effects involve the use of propensity scores, defined as the probability of treatment conditional on baseline covariates. The generalised propensity score (GPS) is an extension of propensity scores which can be applied in settings with multiple treatments [17]. Various effect estimation methods that rely on the GPS have been proposed [17-19] and different methods for estimating the GPS have been considered, including discrete choice models [20] and generalized boosted models [21]. In more recent years, more complex methods for estimating the effects of multiple treatments have been considered, including Bayesian model averaging [22], targeted maximum likelihood estimation [23-24], and joint marginal structural models [25]. Additionally, a compound model was developed that can be used to make treatment decisions when multiple treatment options are available [26].

The aforementioned approaches to estimating multiple treatment effects focus on settings with treatment given at a single time point. However, the effects of treatment strategies that involve sustaining a treatment over time are often of interest. When using longitudinal data with repeated measures of treatment status and covariates over time, estimating effects of sustained treatment strategies requires special methods of analysis to account for time-dependent confounding [27]. Time-dependent confounding arises when confounders of the association between treatment status at a given time and the outcome are influenced by past treatment: adjusting for such confounders may reduce confounding bias, but simultaneously block some of the treatment effect caused by past treatment [28-29]. Robins' Generalised methods (G-methods) are a family of methods which can deal with the issue of time-dependent confounding and include inverse-probability-of-treatment weighted (IPTW) estimation of marginal structural models (IPTW-MSM) [30], g-formula [28] and g-estimation of structural nested models [29]. Another approach, termed "censoring and weighting" [31], is a version of IPTW-MSM which also falls under the umbrella of G-methods. Previous simulation studies [32-34] have compared the performance G-methods for estimating the effects of a sustained treatment strategy involving one binary treatment. However, the methodological literature on approaches to handling time-dependent confounding in settings with multiple treatments is limited. The g-formula [35-36] and IPTW-MSM [37-38] have both been extended to settings with multiple treatments, as has the use of covariate balancing weights as an alternative to IPTW for estimating the parameters of marginal structural models [39]. Whilst different approaches to handling time-dependent confounding and multiple treatments have been proposed, there are currently no studies which compare these approaches.

The aim of this study is to outline how methods that are used to deal with time-dependent confounding can be applied when there are multiple treatments, and to compare them in a simulation study. The remainder of the paper is organised as follows: section 2 describes a motivating example in the context of two treatments used in cystic fibrosis, section 3 describes the methods included in this study, section 4 details the methods and results of our simulation study, section 5 describe the results from applying the methods to the motivating example, and section 6 discusses our findings and concludes.

## 2. Motivating example in cystic fibrosis

Questions involving multiple treatments are particularly relevant in the field of cystic fibrosis (CF). People with CF often report use of ten treatments every day as part of standard care [41], but evidence of treatment efficacy from original randomised controlled trials stems from evaluating individual treatments rather than when used in combination. The complex treatment strategies are associated with high treatment burden, high costs, and increased risk of interactions between medications or adverse drug reactions but are undertaken to optimise health outcomes.

CF is an inherited multisystem disease, with the greatest morbidity affecting the respiratory and digestive system. In the lungs it causes a build-up of thick secretions. Respiratory management includes prompt and aggressive treatment of infection, and promotion of secretion clearance from the airways using nebulised treatments and chest physiotherapy. Nebulised mucoactive treatments make it easier for people with CF to clear these thick secretions. When there is clinical evidence of lung disease, the first choice of mucoactive treatment is dornase alfa (DNase) and approximately 70% of people with CF are currently prescribed DNase as part of their standard care [42]. Depending on patient and clinician preferences, people established on DNase may subsequently follow a number of different treatment strategies. They may continue to take DNase alone, add hypertonic saline to their treatment regimen, or stop treatment with DNase and start hypertonic saline alone (the latter is least common in clinical practice). Existing research provides evidence that DNase used alone, and hypertonic saline used alone, can improve lung function in people with CF [43,44]. However, the combined effects of the two treatments have not been studied in a randomized controlled trial.

In our motivating example, the aim is to compare the effects of treatment strategies involving sustaining combinations of DNase and hypertonic saline on lung function up to 5 years in people with CF using data from the UK CF Registry. The UK CF Registry is a database managed by the Cystic Fibrosis Trust, collecting data on time-invariant variables such as date of birth, genotype and sex, and longitudinal variables that change over time [45]. Longitudinal data are collected at approximately annual intervals when the individual with CF visits the outpatient clinic for a comprehensive review. Longitudinal data includes clinical evaluations (e.g., measurements on weight and lung functions), treatment use over the past year, hospital admissions, culture and microbiology, health complications and outcomes (death or transplants).

In recent work we used UK CF Registry to investigate the effects of combinations of DNase and hypertonic saline on lung function using IPTW-MSM [46]. In this paper we extend this by considering alternative analysis methods. Approvals to use CF Registry data to investigate the effects of DNase and hypertonic saline on clinical outcomes was granted by the CF Registry research committee (Ref: 375). Table 1 describes the research question in more detail, using components of a protocol for a target trial [47]. For the full protocol for the target trial, and the corresponding emulated trial using the UK CF Registry data we refer to [46].

*Table 1: Description of the research question in our motivating example*

| | |
|---|---|
| **Eligibility Criteria** | Include: UK individuals with CF who have been treated with DNase for two years and are aged at least 6 years old. |
| | Exclude: Individuals who have received an organ transplant, been treated with hypertonic saline within the last two years, or are taking mannitol, lumacaftor/ivacaftor or tezacaftor/ivacaftor*. |
| **Treatment Strategies** | 1. Continue Dnase only and do not start hypertonic saline<br>2. Continue Dnase and add hypertonic saline<br>3. Stop Dnase and start hypertonic saline<br>4. Stop Dnase and do not start hypertonic saline<br>Treatment strategy is sustained for the duration of follow-up |
| **Follow-up period** | Up to 5 years |
| **Outcome** | Lung function, measured using forced expiratory volume per 1 second (FEV$_1$%) |
| **Estimand of interest** | Mean differences in FEV$_1$% between treatment strategies at 1-5 years |

*mannitol is another type of mucoactive nebuliser and lumacaftor/ivacaftor and tezacaftor/ivacaftor are CFTR modulators. We excluded individuals on CFTR modulators as we would expect their disease progression to be very different to individuals not taking CFTR modulators. We used data from 2007-2018 which pre-dates the widespread use of CFTR modulators in routine clinical care, so we do not expect large numbers of CFTR modulator users.

## 3. Methods for estimating causal effects with multiple treatments and time-dependent confounding

In this section, we first define the notation, the estimands of interest and the identification assumptions required to estimate them. We then introduce the different analysis approaches for estimation. Five methods are considered: 1) inverse-probability-of-treatment weighting of marginal structural models (IPTW-MSM), 2) censoring and weighting (3) a sequential version of the censoring and weighting approach referred as the "sequential trials" approach , 4) g-formula, and 5) g-estimation of structural nested mean models (SNMM). Our focus throughout is on two treatments, which is the case in the motivating example, and we describe the methods with reference to the example. However, the methods can be straightforwardly extended to three or more treatments.

### 3.1 Notation

Time is denoted by $t$ and is measured in years since baseline (the time of meeting the eligibility criteria). We assume that data on treatments, outcomes and covariates are observed at $T$ discrete time points, where $T = 5$ in the motivating example. Let $A_{it}$ be a time-dependent variable indicating whether individual $i$ is on DNase at time $t$ , and $B_{it}$ be the corresponding indicator for the second treatment, hypertonic saline. Let $Y_{it}$ and $\boldsymbol{L}_{it}$ denote the outcome and a set of covariates for individual $i$ at time $t$, respectively. The set $\boldsymbol{L}_{it}$ may contain time-varying variables ($\boldsymbol{L}'_{it}$) and baseline variables which do not change over time ($\boldsymbol{L}''_{it}$): $\boldsymbol{L}_{it} = \{\boldsymbol{L}'_{it}, \boldsymbol{L}''_{it}\}$. Overbars are used to denote history up to, and including, time $t$; for example: $\bar{A}_{it} = (A_{i0}, A_{i1}, A_{i2}, \dots, A_{it})$. The $i$ subscript is often supressed for simplicity, with the assumption that the random vectors for each individual are drawn from independent and identical distributions. Capital letters have been used to denote random variables; lower-case letters are used to denote possible realisations of those variables. The assumed temporal ordering of the variables is $\{L_0, A_0, B_0, Y_1, L_1, A_1, B_1, Y_2, \ldots, L_{T-1}, A_{T-1}, B_{T-1}, Y_T\}$, though no temporal ordering needs to be assumed for the treatments at a given time point. $Y_{t+1}^{(\bar{a}_t, \bar{b}_t)}$ is the potential outcome for an individual at time $t+1$ had they received treatment history $(\bar{a}_t, \bar{b}_t)$.

## 3.2 Defining causal effects with multiple treatments

Causal estimands are defined as contrasts between functions of potential outcomes under different treatment strategies. For continuous outcomes, we may consider mean differences and for binary outcomes, we may consider risk differences, risk ratios, or odds ratios.

In our motivating example, the outcome is continuous, and we are interested in mean differences. We could be interested in contrasts between $E(Y_{t+1}^{(\bar{a}_t, \bar{b}_t)})$ for any set of values of $(\bar{a}_t, \bar{b}_t)$ for the outcome at any of the time points. . However, not all estimands will be of interest in every study. In our example, we are interested in estimating the following:

$$E\left(Y_{t+1}^{\bar{a}_t=\bar{1},\bar{b}_t=\bar{1}}\right) - E\left(Y_{t+1}^{\bar{a}_t=\bar{1},\bar{b}_t=\bar{0}}\right) \text{ for } t = 0,1,2,3,4 \quad (1)$$

$$E\left(Y_{t+1}^{\bar{a}_t=\bar{0},\bar{b}_t=\bar{1}}\right) - E\left(Y_{t+1}^{\bar{a}_t=\bar{1},\bar{b}_t=\bar{0}}\right) \text{ for } t = 0,1,2,3,4 \quad (2)$$

$$E\left(Y_{t+1}^{\bar{a}_t=\bar{0},\bar{b}_t=\bar{0}}\right) - E\left(Y_{t+1}^{\bar{a}_t=\bar{1},\bar{b}_t=\bar{0}}\right) \text{ for } t = 0,1,2,3,4 \quad (3)$$

Our focus is on people who are already established on DNase, and so estimands (1)-(3) can be interpreted as the effect of: adding hypertonic saline compared to continuing use of DNase alone (1), switching to hypertonic saline compared to continuing use of DNase alone (2) and dropping DNase compared to continuing use of DNase alone (3).

More precisely, when estimating the difference between expected potential outcomes under treatment strategies $(\bar{a}_t, \bar{b}_t)$ and $(\bar{a}'_t, \bar{b}'_t)$, we are estimating the difference in expected potential outcomes between two hypothetical worlds: one in which everyone in the population was observed under treatment strategy $(\bar{a}_t, \bar{b}_t)$ and one in which everyone in the population was observed under treatment strategy $(\bar{a}'_t, \bar{b}'_t)$. A general term for estimands defined this way is the average treatment effect (ATE).

## 3.3 Identification assumptions

Our causal estimands were defined in terms of potential outcomes. To obtain estimates of these from observed data, we need to identify the estimands in terms of the observed data. The assumptions required for identification are no interference, consistency, and sequential conditional exchangeability. Here, we describe these assumptions for settings with two treatments.

***No interference.*** We assume that the treatment strategy observed for individual $i$ at time $t$, $(\bar{A}_{it}, \bar{B}_{it})$, does not influence the potential outcome for individual $j$ at time $s$ ($Y_{js}^{(\bar{a}_{it}, \bar{b}_{it})}$), where $i \neq j$, and for any $s$ and $t$.

***Consistency.*** Consistency states that, for individuals observed under treatment combination $(\bar{a}_t, \bar{b}_t)$, their observed outcome $Y_{t+1}$ is equal to the potential outcome they would have observed had they received the treatments strategy via the hypothetical intervention $(\bar{a}_t, \bar{b}_t)$, $Y_{t+1}^{(\bar{a}_t, \bar{b}_t)}$. For consistency to hold, we require well-defined treatment interventions.

***Conditional exchangeability.*** For a given set of observed covariates $L_t, Y_t$

$$Y_{t+1}^{(\bar{a}_t, \bar{b}_t)} \perp\!\!\!\perp (\bar{A}_t, \bar{B}_t) | \bar{L}_t, \bar{Y}_t, \forall (\bar{a}_t, \bar{b}_t) \text{ and } \forall t$$

Conditional exchangeability states that the observed treatment combination is independent of each of the potential outcomes at each time point, conditional on $\bar{C}_t$. This is often known as the "no unmeasured confounders" assumption.

No interference, consistency and conditional exchangeability are the identification assumptions. For estimation, we also require the positivity assumption.

**Positivity.** $0 < P(\bar{A}_t = \bar{a}_t, \bar{B}_t = \bar{b}_t | \bar{C}_t) < 1 \; \forall \; \bar{C}_t$ and $(\bar{a}_t, \bar{b}_t)$.

Positivity states that, for all possible values of $\bar{C}_t$, it must not be inevitable or impossible to observe any of the treatment combinations.

### 3.4 Inverse-probability-of-treatment weighting of marginal structural models

Inverse-probability-of-treatment weighting of marginal structural models (IPTW-MSM) involves assigning weights to each individual based on their propensity score [30]. The propensity score at time $t$ is defined as the probability of individuals receiving the treatment combination they received at that time, conditional on treatment and covariate history up to that time. We define a new categorical variable denoting which treatment combination is observed for an individual at time $t$:

$$Z_t = \begin{cases} 0 \text{ if } A_t = 0 \text{ and } B_t = 0 \\ 1 \text{ if } A_t = 1 \text{ and } B_t = 0 \\ 2 \text{ if } A_t = 0 \text{ and } B_t = 1 \\ 3 \text{ if } A_t = 1 \text{ and } B_t = 1 \end{cases} \quad (4)$$

Assuming that $L$ is a set of variables for which conditional exchangeability holds, the propensity score is defined as:

$$PS_t = \Pr(Z_t = z_t | \bar{Z}_{t-1} = \bar{z}_{t-1}, \bar{L}_t = \bar{l}_t, \bar{Y}_{t-1} = \bar{y}_{t-1}) \quad (5)$$

These can be estimated using multinomial regression with a categorical variable denoting treatment combination as the outcome and treatment and covariate histories as the predictors. Alternatively, $\Pr(A_t = a_t | \bar{A}_{t-1} = \bar{a}_{t-1}, \bar{B}_{t-1} = \bar{b}_{t-1}, \bar{L}_t = \bar{l}_t, \bar{Y}_{t-1} = \bar{y}_{t-1})$ and $\Pr(B_t = b_t | \bar{B}_{t-1} = \bar{b}_{t-1}, \bar{A}_{t-1} = \bar{a}_{t-1}, \bar{L}_t = \bar{l}_t, \bar{Y}_{t-1} = \bar{y}_{t-1})$ could be estimated via separate logistic regression models, and used to obtain estimates of the probability of each treatment combination by multiplying. The latter approach would be increasingly necessary with more than two treatments, as the number of combinations then becomes large, but make the additional assumption that the treatments are independent conditional on the covariates included in the models. Using multinomial regression also imposes the constraint that all predicted probabilities sum to 1, whereas this restriction is not possible when using multiple logistic regression models.

Once propensity scores are estimated, the weight at each time is then equal to the cumulative product of estimated propensity scores up to that time:

$$W_t = \prod_{j=1}^{t} \frac{1}{PS_j} \quad (6)$$

This approach can lead to extreme weights, which result in unstable effect estimates. To overcome this issue, stabilized weights can be used. A general formula for the stabilised inverse-probability-of-treatment weights is given by:

$$SW_t = \prod_{j=1}^{t} \frac{\Pr(Z_j = z_j | \bar{Z}_{j-1} = \bar{z}_{j-1})}{PS_j} \quad (7)$$

A marginal structural model (MSM) is then fitted using weighted regression. The MSM specifies how the outcome at a given time depends on treatment history up to that time. For example, we may assume that the outcome depends on the duration each treatment, and include their interaction:

$$E\left(Y_{t+1}^{(\bar{a}_t, \bar{b}_t)}\right) = \alpha_{0t} + \alpha_1 \sum_{j=0}^{t} a_j + \alpha_2 \sum_{j=0}^{t} b_j + \alpha_3 \sum_{j=0}^{t} a_j b_j \quad (8)$$

This model assumes that each additional year of treatment has the same effect on the outcome, regardless of how long an individual has been on treatment. Alternatively, we can allow the effects of treatments at different time points to vary. There are various ways to do this, and one example is to use an MSM of the form:

$$E\left(Y_{t+1}^{(\bar{a}_t,\bar{b}_t)}\right) = \beta_{0t} + \sum_{j=0}^{t} \beta_{1j} a_j + \sum_{j=0}^{t} \beta_{2j} b_j + \sum_{j=0}^{t} \beta_{3j} a_j b_j \qquad (9)$$

To estimate causal effects, we use the results from the fitted MSM to estimate expected potential outcomes under different treatment strategies and then calculate the relevant measure of association. When using stabilised weights, including time-invariant confounders in the numerator can help stabilisation [48]. However, if this is done, the covariates included in the numerator must also be included in the MSM.

The sandwich estimator of the standard error is often invalid in this setting as it does not consider the fact that the weights have been estimated, and this can lead to overestimation of the variance and therefore conservative confidence intervals [49]. Instead, bootstrapping can be used to obtain valid standard error estimates.

### 3.5 Censoring and weighting approaches

The censoring and weighting approach is closely related to the IPTW-MSM approach. In the censoring and weighting approach, individuals are artificially censored (i.e., their follow-up is cut-off) when they deviate from their baseline treatment status [31]. This extends directly to the setting with multiple treatments, with individuals being artificially censored when they deviate from their treatment combination at baseline. Inverse-probability-of-censoring weights (IPCW) are used to account for the selection bias induced by artificial censoring. The IPCW at a given time is the inverse of the probability of being observed (i.e., remaining uncensored) at time $t$. Because censoring is determined by treatment status, these weights are equivalent to those in equation (3) for people who remain uncensored at a given time. Like IPTW, weight stabilisation can be used to improve precision.

The expectation $E\left(Y_{t+1}^{(\bar{a}_t,\bar{b}_t)}\right)$ can be estimated non-parametrically by the weighted mean of $Y_{t+1}$ among those observed to follow strategy $(\bar{a}_t, \bar{b}_t)$. In the motivating example our interest lies only in four strategies: $(\bar{a}_t = 0, \bar{b}_t = 0)$, $(\bar{a}_t = 0, \bar{b}_t = 1)$, $(\bar{a}_t = 1, \bar{b}_t = 0)$, $(\bar{a}_t = 1, \bar{b}_t = 1)$. We note that under these strategies, after implementing the artificial censoring, in the remaining data, we have $a_t = a_0$ and $b_t = b_0$ for all individuals at all $t$. Alternatively MSMs could be fitted to the modified data by weighted estimation, and there are different ways in which this could be done. For example, we could fit an MSM that assumes the outcome depends only on baseline treatment, which here is equivalent to current treatment, but not on treatment duration:

$$E\left(Y_{t+1}^{(\bar{a}_t,\bar{b}_t)}\right) = \alpha_{0t} + \alpha_1 a_0 + \alpha_2 b_0 + \alpha_3 a_0 b_0 \qquad (10)$$

Or we could allow for the treatment effects to vary over time, for example using

$$E\left(Y_{t+1}^{(\bar{a}_t,\bar{b}_t)}\right) = \alpha_{0t} + \alpha_{1t} a_0 + \alpha_{2t} b_0 + \alpha_{3t} a_0 b_0 + \alpha_{4t} a_0 t + \alpha_{5t} b_0 t + \alpha_{6t} a_0 b_0 t \qquad (11)$$

The censoring and weighting approach focuses on the treatment strategies of interest, rather than all possible strategies. When there are two treatments, model (10) is equivalent to including a categorical variable representing all four possible treatment combinations. However, when there are more than two treatments, an MSM may only represent a subset of treatment combinations (those of interest) and only data from the individuals observed on those combinations is used to fit the model. Conversely, in an the IPTW-MSM analysis, the MSM is fitted using data on all individuals at all time-points at which they are observed. The IPTW-MSM approach typically makes assumptions that borrow information across longitudinal treatment patterns, i.e., by using an unsaturated MSM, even though the focus is typically on contrasts between outcomes under a subset of the possible treatment patterns. The censoring and weighting approach may also make such assumptions via an MSM, such as in (10) or (11), but will borrow information across longitudinal treatment patterns to a lesser extent.

A limitation of the censoring and weighting approach is that it may involve censoring many individuals, which will decrease the precision and this is likely to be much more of an issue in the multiple treatments setting. Variances can be estimated using the bootstrap approach.

## 3.6 Sequential trials approach

The sequential trials method is an extended version of the censoring and weighting approach, which aims to improve efficiency by making use of more of the data [31] by applying the censoring and weighting from a series of new time origins. In the sequential trials method, we view the observational data as a sequence of nonrandomised "nested trials". A new "trial" is started at each time point $0, \ldots, T - 1$, and any individuals who meet the inclusion and exclusion criteria at that time are included in the trial. In our context the inclusion criteria for the trial starting at time $t$ include that the individual has not yet initiate either treatment of interest, $\bar{A}_{t-1} = 0, \bar{B}_{t-1} = 0$. The subset of individuals considered for inclusion in the trial starting at time $t + 1$ is (approximately – dependent on other inclusion/exclusion criteria) nested within in the subset of individuals considered for inclusion in the prior trial starting at time $t$. Artificial censoring is implemented within each trial when individuals deviate from their treatment combination at the start of the trial, as in the censoring and weighting approach. .

For the trial starting at time $k$ ($k = 0, \ldots, T - 1$) we let $L_{k,t}, A_{k,t}, B_{k,t}$ denote covariates and treatment statuses at time $t$ after the start of the trial, and $Y_{k,t+1}$ the outcome at time $t + 1$ after the start of the trial ($t = 0, \ldots, T_1$). We let $Y_{k,t+1}^{(\bar{a}_t, \bar{b}_t)}$ denote the potential outcome at time $t + 1$ after the start of trial $k$ if an individual were to follow treatment strategy $(\bar{a}_t, \bar{b}_t)$ up to time $t$ from the start of the trial. The expectation $E\left(Y_{k,t+1}^{(\bar{a}_t, \bar{b}_t)}\right)$ could be estimated non-parametrically by the weighted mean of $Y_{k,t+1}$ among those observed to follow strategy $(\bar{a}_t, \bar{b}_t)$, separately for each trial $k$. However, a combined analysis across trials is typically performed by specifying a marginal structural model for $E\left(Y_{k,t+1}^{(\bar{a}_t, \bar{b}_t)}\right)$. For example, corresponding to the MSM in (10),

$$E\left(Y_{k,t+1}^{(\bar{a}_t, \bar{b}_t)}\right) = \alpha_{0kt} + \alpha_1 a_0 + \alpha_2 b_0 + \alpha_3 a_0 b_0 \tag{12}$$

As in this example MSM, coefficients for treatment are typically considered to be common across trials and the model is fitted using a weighted regression combined across trials. The weights can be estimated in a pooled way across trials or separately by trial. Care needs to be taken to consider whether the sequential trials analysis targets the same marginal estimand as other approaches [50]. In the combined data across trials, the distribution of baseline characteristics $L_{k,0}$ (for $k = 0, \ldots, T - 1$ combined) is likely to differ from the distribution of characteristics $L_0$ of the population at the overall time 0, which is the population to which the marginal estimands targeted by the other approaches to estimating contrasts between $E\left(Y_{t+1}^{(\bar{a}_t, \bar{b}_t)}\right)$ under different treatment strategies refer. Under linear MSMs in which there are assumed to be no interactions between treatment and baseline covariates, such as in model (12), the marginal estimates $\hat{E}\left(Y_{k,t+1}^{(\bar{a}_t, \bar{b}_t)}\right) - \hat{E}\left(Y_{k,t+1}^{(\bar{a}'_t, \bar{b}'_t)}\right)$ will target the same marginal estimand as $\hat{E}\left(Y_{t+1}^{(\bar{a}_t, \bar{b}_t)}\right) - \hat{E}\left(Y_{t+1}^{(\bar{a}'_t, \bar{b}'_t)}\right)$. More generally however it may be necessary to take additional steps to ensure that the sequential trials analysis targets the intended marginal estimand – for example if there are interactions between treatment and baseline covariates in the MSM or if the outcome model were a logistic regression model. This could be by specifying an MSM conditional on baseline covariates $E\left(Y_{k,t+1}^{(\bar{a}_t, \bar{b}_t)} | L_{k,0}\right)$ and then standardising to the population of interest [50].

A robust variance estimator which accounts for within-person correlation can be used [31]. This is important as many individuals will be included in multiple "trials". Alternatively, bootstrapping can be used to estimate the variance, as in the above methods.

## 3.7 The G-formula

The g-formula is a generalisation of standardisation to settings where treatment and covariates vary over time [35]. It is given by:

$$E\left(Y_{t+1}^{(\bar{a}_t,\bar{b}_t)}\right) = \int E\{[Y_{t+1}|\bar{A}_t = \bar{a}_t, \bar{B}_t = \bar{b}_t, \bar{L}_t = \bar{l}_t, \bar{Y}_t = \bar{y}_t]\} \prod_{j=0}^{t} f(l_j|\bar{a}_{j-1}, \bar{b}_{j-1}, \bar{l}_{j-1}, \bar{y}_{j-1}) d\bar{l} \quad (13)$$

where the integral is over all possible values $\bar{l}_t$.

To implement the g-formula, we must specify and fit models for $E[Y_{t+1}|\bar{A}_t = \bar{a}_t, \bar{B}_t = \bar{b}_t, \bar{L}_t = \bar{l}, \bar{Y}_t = \bar{y}_t]$ and $P(l_j|\bar{a}_{j-1}, \bar{b}_{j-1}, \bar{l}_{j-1}, \bar{y}_{j-1})$. This involves postulating models for $Y_{t+1}$ conditional on $\bar{A}_t, \bar{B}_t, \bar{L}_t$ and $\bar{Y}_t$, and models for $L_t$ conditional on $\bar{A}_{t-1}, \bar{B}_{t-1}, \bar{L}_{t-1}$ and $\bar{Y}_{t-1}$. The models could be fitted separately at each time point, or could be fitted in a pooled way across time points. With multiple time points and multi-dimensional $L$, implementing the g-formula analytically becomes infeasible. To overcome this difficulty, an algorithm involving Monte Carlo simulations can be used [51]. Using the observed baseline data $L_0$ as a starting point, the algorithm simulates complete datasets where the treatment strategy is predetermined by the analyst, and time-varying covariates and outcomes are iteratively drawn from the conditional densities previously specified. Expected potential outcomes for each predetermined treatment strategy are then computed as the mean end of follow-up outcome across the simulated data. Treatment effect estimates are obtained as the difference in expected potential outcomes. Standard errors can be obtained using either the delta method, or bootstrapping [27].

### 3.8 G-estimation of structural nested models

G-estimation is used to estimate structural nested mean models (SNMM) [52-53]. A SNMM is a set of sub-models, one for each time horizon at which we wish to estimate the treatment effect. At a given time, the SNMM is a model for the effect of treatment versus no treatment at that time, in the hypothetical case where treatment is set to zero after that time (where 0 indicates no treatment). In our case of two treatments, the SNMM is a model for the effect of some combination of both treatments versus neither treatment at that time, in the hypothetical case where both treatments are set to zero after that time. For example, the SNMM defined by:

$$E\left(Y_s^{((\bar{a}_t,0),(\bar{b}_t,0))} - Y_s^{((\bar{a}_{t-1},0),(\bar{b}_{t-1},0))}\middle|\bar{a}_{t-1},\bar{b}_{t-1},l_t\right) = \varphi_A a_t + \varphi_B b_t + \varphi_{AB} a_t b_t \quad (14)$$

postulates that the treatment combination $(a_t, b_t)$ has the same effect on the outcome $Y_s$ for all $s = 1, ..., T$ (where $T$ is the maximum follow-up year) and $t < s$. This SNMM model is the natural analogue to the MSM defined in equation (8).

A general form for a linear SNMM for settings with two treatments is:

$$E\left(Y_s^{((\bar{a}_t,0),(\bar{b}_t,0))} - Y_s^{((\bar{a}_{t-1},0),(\bar{b}_{t-1},0))}\middle|\bar{a}_{t-1},\bar{b}_{t-1},l_t\right) = \boldsymbol{\varphi}_A \boldsymbol{w}_{A,st} a_t + \boldsymbol{\varphi}_B \boldsymbol{w}_{B,st} b_t + \boldsymbol{\varphi}_{AB} \boldsymbol{w}_{AB,st} a_t b_t \quad (15)$$

for all $s = 1, ..., T$ and $t < s$. $\boldsymbol{w}_{A,st}$, $\boldsymbol{w}_{B,st}$ and $\boldsymbol{w}_{AB,st}$ are vectors that may include functions of $l_t$ and $t$, allowing for effect modification by covariate history, or effects that vary with time. $\boldsymbol{\varphi}_A$, and $\boldsymbol{\varphi}_B$ are vectors whose elements sum to the effect of $(A_t = 1, B_t = 0)$ and $(A_t = 0, B_t = 1)$ on $Y_s$, and $\boldsymbol{\varphi}_{AB,st}$ is a vector whose elements sum to the additional effect caused by the interaction between $A_t$ and $B_t$.

To estimate $\boldsymbol{\varphi}_A, \boldsymbol{\varphi}_B$ and $\boldsymbol{\varphi}_{AB}$, we use the approach recommended by Vansteelandt et al [52]. Firstly, we fit a model for $Y_s$ conditional on $A_{s-1}, B_{s-1}$, and their interaction, treatment and confounder history $(\bar{A}_{s-2}, \bar{B}_{s-2}, \bar{L}_{s-1})$ and the estimated propensity score $(\widehat{PS}_{s-1})$. This model is fitted for all $s = 1, ..., T$ combined and provides initial estimates of $\widehat{\boldsymbol{\varphi}}_A, \widehat{\boldsymbol{\varphi}}_B$ and $\widehat{\boldsymbol{\varphi}}_{AB}$. We then apply the following two steps recursively:

Step 1: Define $H_{st}$ as the potential outcome $Y_s$ given treatment is set to its history up to time $t$ and to 0 thereafter, i.e., $((\bar{a}_t, 0), (\bar{b}_t, 0))$. $H_{st}$ is estimated as follows:

$$\widehat{H}_{st} = Y_s - \sum_{u=t+1}^{s}(\widehat{\boldsymbol{\varphi}}_A \boldsymbol{W}_{A,u} A_u + \widehat{\boldsymbol{\varphi}}_B \boldsymbol{W}_{B,u} B_u + \widehat{\boldsymbol{\varphi}}_{AB} \boldsymbol{W}_{AB,u} A_u B_u) \text{ for } t \leq s \quad (16)$$

Step 2: Update $\hat{\varphi}$ by fitting a model for $\hat{H}_{st}$ conditional on $A_{t-1}, B_{t-1}$, and their interaction, treatment and confounder history ($\bar{A}_{t-2}, \bar{B}_{t-2}, \bar{L}_{t-1}$) and the estimated propensity score ($\widehat{PS}_{s-1}$).

Steps 1 and 2 are first applied for $t = s - 1$. They are then repeated using the updated $\hat{\varphi}$ to re-predict existing potential outcomes and predict further potential outcomes with an additional time period between the treatment and the outcome. This process is repeated until $H_{s(s-T)}$ is estimated and used to obtain a final estimate of $\varphi$. Standard errors may be estimated using a sandwich estimator [27] or the bootstrap method.

### 3.9 Comparison of methods

IPTW-MSM is the most widely used method in the applied literature [54]. Authors have suggested that possible reasons for this are the perceived simplicity of this approach [52,54]. A disadvantage of IPTW-MSM is that it is inefficient compared to both the g-formula [27] and g-estimation [52]. This is due to extreme weights which lead to increased variance in treatment effect estimates, and this issue is often exacerbated as the number of confounders increase [27]. In the setting of estimating combined effects of multiple treatments the problem of extreme weights is expected to be even greater. Moreover, IPTW-MSM is susceptible to bias when there is a strong association between confounder and treatment [52].

The g-formula, while more efficient than IPTW-MSM, is more susceptible to bias due to model misspecification. IPTW-MSM only requires postulation of the models used for the weights and the MSM, whereas g-formula requires postulation of models for all time-varying confounders. As the number of confounders increase, the inefficiency of IPTW-MSM increases, as does the risk of model misspecification in the g-formula. Another limitation of the g-formula is that some degree of model misspecification is guaranteed when there is treatment-confounder feedback, and the treatment has no causal effect on the outcome [55]. A consequence of this is that, given enough data, we would reject the causal null hypothesis even when it is true. This is often referred to as the *g-null paradox* [56].

G-estimation is typically more efficient than IPTW-MSM and requires fewer parametric assumptions than the g-formula. For this reason, it is often considered a good compromise for the bias-variance trade-off. Another advantage of g-estimation is that it can incorporate effect modification by time-varying covariates. However, g-estimation is not well suited to binary outcomes, although Vansteelandt [57] has suggested models to overcome this limitation.

Censoring and weighting, and the sequential trials approach, are increasingly used in the applied literature. Like IPTW-MSM, they are relatively easy to understand and implement, and only require postulation of treatment and outcome models, so are less susceptible to model misspecification bias than the g-formula. However, these approaches may be inefficient over longer follow-up periods in situations where many individuals switch treatment combinations and are therefore censored. In the multiple treatments setting we expect more individuals to be artificially censored under these approaches, resulting in increased inefficiency.

## 4. Simulation study

The aim of this stimulation study is to compare the performance of the five analysis methods outlined above for estimating the effects of two treatments in combination in settings with time-dependent confounding. Section 4.1 describes the study design and analysis plan, and this section follows the reporting guidance of Morris et al [58]. Section 4.2 presents the results.

### 4.1. Simulated Data

*4.1.1 Data Generation*

We simulated nine different scenarios, each with a sample size of 10,000. All scenarios included two binary treatment variables, $A$, and $B$, a continuous outcome, $Y$, and a continuous covariate, $L$. All variables were time-varying across six time points, $t = 0, ..., 5$. Baseline refers to the time when $t = 0$. Data were simulated sequentially, and Figure 1 shows a DAG depicting the causal pathways between each of our variables over time. Figure 1 shows that both $L_t$ and $Y_t$ are time-varying confounders for the causal relationships between $(A_t, B_t)$ and $Y_{t+1}$. The formulae used to generate data in all scenarios are provided in the supplementary file (see section S.1.1).

Scenario 1 (S1) was the simplest scenario and is used as a comparison reference for the more complex scenarios. Scenarios 2-5 varied the way the treatment, confounders, or outcome were generated. (Table 2). They were designed to assess the effect of: reducing the proportion of individuals using one treatment (S2); increasing the strength of association between $L_t$ and $Y_{t+1}$ (S3); increasing the strength of association between the $L_t$ and one of the treatments (S4); and introducing statistical dependence between probability of receiving treatments $A$ and $B$ (S5). Scenarios 6-9 varied the treatment effect (Table 2). Variations included: reducing the causal effect of one treatment to 0 (S6); adding an interaction effect between treatments (S7); adding effects of past treatment use on current outcomes (S8); and introducing a treatment effect which decreases over time (S9).

*Table 2: Description of scenarios 1-9 and changes to the DAG between scenarios*

| Scenario | Description | Changes to the DAG (Figure 1) |
|---|---|---|
| S1 | Baseline scenario | None |
| **Changes to the way treatment, confounders, or outcome were generated** | | |
| S2 | Lower prevalence of $A$ | None |
| S3 | Stronger association between $L_t$ and $Y_{t+1}$ | None |
| S4 | Stronger association between $L_t$ and $A_t$ | None |
| S5 | Introduce dependence between $A$ and $B$: $A_t$ depends on $B_{t-1}$ and vice versa | Add $A_{t-1} \to B_t$ and $B_{t-1} \to A_t$ |
| **Changes to simulated treatment effects** | | |
| S6 | $A$ has no causal effect on $Y$ | Remove $A_t \to Y_{t+1}$ $\forall\, t$ |
| S7 | $A \times B$ has a causal effect on $Y$ | None |
| S8 | $Y_{t+1}$ depends on entire treatment history | Add arrows between each element of $\bar{A}_t$ and $Y_{t+1}$ |
| S9 | The size of the causal effect that $A_t$ has on $Y_{t+1}$ decreases as the cumulative years on $A$ increases | None |

### 4.1.2 Estimands of interest

We are interested in static sustained treatment regimes based on $A$ only, $B$ only, and $A$ and $B$ in combination. The estimands of interest were all possible comparisons between the treatment combinations at all time points:

$$ATE_{t+1}^{\{A-0\}} = E\left(Y_{t+1}^{\bar{A}_t=\bar{1}_t, \bar{B}_t=\bar{0}_t}\right) - E\left(Y_{t+1}^{\bar{A}_t=\bar{0}_t, \bar{B}_t=\bar{0}_t}\right) \text{ for } t = 0,1,2,3,4$$
$$ATE_{t+1}^{\{B-0\}} = E\left(Y_{t+1}^{\bar{A}_t=\bar{0}_t, \bar{B}_t=\bar{1}_t}\right) - E\left(Y_{t+1}^{\bar{A}_t=\bar{0}_t, \bar{B}_t=\bar{0}_t}\right) \text{ for } t = 0,1,2,3,4$$
$$ATE_{t+1}^{\{A-B\}} = E\left(Y_{t+1}^{\bar{A}_t=\bar{1}_t, \bar{B}_t=\bar{0}_t}\right) - E\left(Y_{t+1}^{\bar{A}_t=\bar{0}_t, \bar{B}_t=\bar{1}_t}\right) \text{ for } t = 0,1,2,3,4$$
$$ATE_{t+1}^{\{AB-0\}} = E\left(Y_{t+1}^{\bar{A}_t=\bar{1}_t, \bar{B}_t=\bar{1}_t}\right) - E\left(Y_{t+1}^{\bar{A}_t=\bar{0}_t, \bar{B}_t=\bar{0}_t}\right) \text{ for } t = 0,1,2,3,4$$
$$ATE_{t+1}^{\{AB-A\}} = E\left(Y_{t+1}^{\bar{A}_t=\bar{1}_t, \bar{B}_t=\bar{1}_t}\right) - E\left(Y_{t+1}^{\bar{A}_t=\bar{1}_t, \bar{B}_t=\bar{0}_t}\right) \text{ for } t = 0,1,2,3,4$$
$$ATE_{t+1}^{\{AB-B\}} = E\left(Y_{t+1}^{\bar{A}_t=\bar{1}_t, \bar{B}_t=\bar{1}_t}\right) - E\left(Y_{t+1}^{\bar{A}_t=\bar{0}_t, \bar{B}_t=\bar{1}_t}\right) \text{ for } t = 0,1,2,3,4$$

The six possible treatment combination pairwise comparisons and five time points give a total of 30 estimands. For the main results, we focus on one set of comparisons: $ATE_{t+1}^{\{AB-B\}}$ for $t = 0,1,2,3,4$.

### 4.1.3 Analysis methods and implementation

The performances of five methods are compared: 1) IPTW of MSMs, 2) censoring and weighting, 3) sequential trials, 4) g-formula, and 5) g-estimation of SNMMs. All methods involved estimating propensity scores, i.e., the probability of treatment received, conditional on treatment and confounder history. To estimate propensity scores, we used multinomial regression:

$$\ln\left(\frac{P(Z_t=z_t)}{P(Z_t=0)}\right) = \beta_{0,z} + \sum_{c=1}^{3}\beta_{cz}I(z_{t-1}=c) + \beta_{B,z}Y_0 + \beta_{Y,z}Y_{t-1} + \beta_{L,z}L_t, \text{ for } z=1,2,3 \quad (17)$$

Where $Y_0$ is the outcome measured at baseline, and $Z_t$ denotes the treatment combination at time $t$ (equation (4)). The propensity scores were used to estimate weights in the IPTW, censoring, and sequential trials methods (see equation S.12 in the supplementary file).

For IPTW, the following marginal structural model was fitted in the weighted population:

$$E(Y_{t+1}^{\bar{z}_t}) = \alpha_0 + \sum_{j=1}^{t}\sum_{c=1}^{3}\alpha_{cj}I(z_j=c) + \alpha_Y Y_0 + \varepsilon_{it} \quad (18)$$

For a fair comparison across scenarios, the same model (model 18) was used in all scenarios, despite being over-specified in some scenarios where the data generating mechanism was simpler. We have included $Y_0$ as a baseline confounder, as this was included in the numerator of the stabilised weights. The intercept here is not time dependent as that wasn't necessary given the way the data were simulated.

For censoring and weighting, individuals were censored when they deviated from the treatment combination they were on at time $t=0$. The uncensored population was weighted (using weights as defined in equation S.12 in the supplementary file) and model (18) was then fitted in the censored and weighted population. Model 18 was used here to encourage a fair comparison between this approach and IPTW.

The sequential trials approach included five "trials" with baseline at times $t=0,1,2,3,4$. At time $t$, individuals were eligible for inclusion if they had no history of treatment, i.e., $\bar{z}_t=\bar{0}$. All trials were then pooled into one dataset for analysis. The censoring and weighting approach was used for analysis with one modification: an indicator for "trial" was included in the propensity score model and the marginal structural model.

For the g-formula, the following model was used to estimate outcomes under each treatment strategy:

$$E(Y_{t+1}|\bar{Z}_t, Y_t, L_t) = \rho_0 + \rho_L L_t + \sum_{j=1}^{t}\sum_{c=1}^{3}\rho_{cj}I(z_j=c) + \rho_Y Y_t + \varepsilon_{it} \quad (19)$$

The g-formula also requires specification of models for time-varying confounders, and these are provided in the supplementary file (see equations S.6 and S.7).

For g-estimation, we estimated the SNMM defined in (15) with $w_{A,st}=w_{B,st}=w_{AB,st}=(1,1,1,1,1)^T$. The following outcome model was used to obtain an initial estimate of $\varphi$:

$$E(Y_T|\bar{Z}_{T-1}, Y_{T-1}, L_{T-1}, PS_{T-1}) \quad (20)$$
$$= \delta_0 + \sum_{c=1}^{3}\varphi_c I(z_{T-1}=c) + \sum_{j=0}^{T-2}\sum_{c=1}^{3}\delta_{cj}I(z_j=c) + \delta_L L_{T-1} + \delta_Y Y_{T-1}$$
$$+ \delta_{PS}PS_{T-1}$$

where $PS_{T-1}$ is the estimated propensity score at time $T-1$. We then estimated $H_{st}$ using $\hat{\varphi}$ (equation 16) and used $\hat{H}_{st}$ to update the estimate of $\varphi$ by fitting model (21) in the population. This process was repeated for all $s=1,\ldots,T-1$ and $t\leq s$, until a final estimate of $\varphi$ was obtained.

$$E(H_{st}|\bar{Z}_t, Y_t, L_t, PS_t) \quad (21)$$
$$= \delta_0 + \sum_{c=1}^{3}\varphi_c I(z_t=c) + \sum_{j=0}^{t-1}\sum_{c=1}^{3}\delta_{cj}I(z_j=c) + \delta_L L_t + \delta_Y Y_t + \delta_{PS}PS_t$$

In all methods, outcome models were fitted using multivariable linear regression.

*4.1.4 Performance measures*

In each scenario, the analysis methods were compared in terms of bias and empirical standard error. Monte Carlo standard errors for both performance measures were also estimated. These quantify the uncertainty in our estimate of the performance measure due to the finite number of simulated data sets [58]. In our study, 1000 datasets were simulated for each scenario ($n_{sim} = 1000$). Monte Carlo standard errors were used to obtain 95% confidence intervals for the performance measures. Table 3 defines the equations used to estimate the bias, empirical standard error and their Monte Carlo standard errors.

*Table 3: Performance measures: estimates and Monte Carlo standard errors (SE). $n_{sim}$ denotes the number of simulated datasets; $\theta$ denotes the true value of the estimand; $\hat{\theta}_j$ denotes the estimate of $\theta$ obtained in the jth dataset and $\bar{\theta}$ denotes the mean of $\hat{\theta}_j$ across the $n_{sim}$ datasets.*

| Performance Measure | Estimate | Monte Carlo SE of Estimate |
|---|---|---|
| Bias | $\theta - \bar{\theta}$ | $\sqrt{\dfrac{1}{n_{sim}(n_{sim}-1)}\sum_{j=1}^{n_{sim}}(\hat{\theta}_j - \bar{\theta})^2}$ |
| Empirical standard error (EmpSE) | $\sqrt{\dfrac{1}{n_{sim}-1}\sum_{j=1}^{n_{sim}}(\hat{\theta}_j - \bar{\theta})^2}$ | $\dfrac{\widehat{EmpSE}}{\sqrt{2((n_{sim}-1))}}$ |

The true treatment effects were obtained by generating longitudinal data for a large "randomised controlled trial", where the relationships between the variables are similar as those in the simulation study, except $A_t$ and $B_t$ are not affected by $L_t$ or $Y_{t-1}$. Instead, they are set to the fixed value determined by the desired treatment combination. For each treatment combination of interest, we generated randomised controlled trials with 1,000,000 individuals and true treatment effects were obtained by taking the difference between the expected value of the relevant outcomes.

### 4.2 Results

We present the results obtained when estimating the effect of treatment combination $A$ and $B$ compared to $B$ (i.e., $ATE_{t+1}^{\{AB-B\}}$), for $t = 0,1,2,3,4$, using IPTW-MSM, censoring and weighting, sequential trials, g-formula and g-estimation. Results obtained for the alternative estimands are presented in the supplementary file (section S.1.4).

*4.2.1 Bias*

Figure 2 shows the bias, and 95% confidence intervals for bias, obtained when estimating the effect of $A$ and $B$ versus $B$ only for all time points and all scenarios.

All methods resulted in approximately unbiased estimates of the treatment effect in scenarios 1-8. Most of the 95% confidence intervals for bias contained 0; in cases where 0 fell outside of the confidence interval, the estimated bias was negligible (Figure 2).

In scenario 9, all methods obtained biased effect estimates except for censoring and weighting, and sequential trials. In this scenario, the effect of $A$ on $Y$ decreased over time and this was simulated by including the interaction between $A$ and the cumulative years treated with $A$ in the outcome model, with a negative coefficient. Consequently, when $t = 3$, for example, the effect of $A_3$ on $Y_4$ would be smaller for individuals who had been on treatment $A$ for 3 years, compared to an individual who initiated treatment $A$ 1 or 2 years ago. When estimating the effect of $A_3$ on $Y_4$ using IPTW, g-formula, or g-estimation, the outcome models use data from individuals who have been treatment $A$ for 1, 2 or 3 years, incorrectly assuming that the effect is the same, regardless of treatment duration. When estimating the effect of $A_3$ on $Y_4$ using censoring and weighting, or sequential trials, only data from individuals who have been on treatment $A$ for 3 years will be used, as individuals who deviate from their original treatment combination are censored. Consequently, the latter two methods obtain unbiased estimates.

*4.2.2 Empirical Standard Error*

Figure 3 shows the empirical standard error, and corresponding 95% confidence intervals, obtained when estimating the effect of $A$ and $B$ versus $B$ only for all time points and all scenarios.

For all methods and in all scenarios, the empirical standard error tends to increase by time. The rate of increase is greatest for the results obtained by censoring and weighting method, and sequential trials. This is because these methods censor individuals when they deviate from their original treatment combination and so the sample size decreases over time.

The efficiency of censoring and weighting, and sequential trials, depends on the proportion of individuals who remain uncensored. In scenario 1, on average (across the 1000 datasets), only 3% of individuals who started on $B$ only, remained on $B$ only for 5 years, and 40% of individuals who started on $A$ and $B$ remained on $A$ and $B$ for 5 years. The equivalent percentages for scenario 2 are 60% and 57%. Therefore, compared to scenario 1, less people were censored in scenario 2 and hence the standard errors obtained using censoring and weighting, or sequential trials, are smaller. The standard errors obtained using sequential trials approach are marginally smaller than those obtained using the censoring and weighting approach for most time points. This is as expected since the sequential trials approach gains efficiency by including individuals in the analysis more than once.

IPTW and g-estimation perform very similarly in most scenarios in terms of efficiency and tend to obtain smaller standard errors than g-formula. In scenario 4 however, the standard error in IPTW is consistently greater than g-estimation. In scenario 4 there was a stronger association between the covariate ($L$) and treatment ($A$), leading to large imbalances in the distribution of the confounder between the group of people taking $A$ and the group not taking $A$. These imbalances led to larger weights which can obtain less stable estimates, hence the increase in variance.

## 5. Motivating example: data and results

We use the motivating example introduced in section 2 to illustrate the application of the five different analysis approaches to addressing a clinical question using real data. The question involves comparisons of multiple treatment strategies involving DNase and hypertonic saline and compares their effects on lung function in people with cystic fibrosis (CF) (Table 1).

### 5.1 Data source and methods

We used 2007-2018 data from the UK CF Registry [59]. We included individuals aged 6 years or older, who have been prescribed DNase for two consecutive years between 1st January 2007 and 31st December 2017. Exclusions occur if they had been prescribed hypertonic saline within the two years prior to study entry, prescribed mannitol, lumacaftor/ivacaftor, or tezacaftor/ivacaftor, or have ever received an organ transplant. Time 0 is the first year inclusion and exclusion criteria are met. Individuals were followed for up to 5 years. Some individuals had less than 5 years follow-up due to administrative end of follow-up, death, or transplant and these individuals were censored. The supplementary file provides further details on how the methods were applied, which variables were treated as time-invariant and time-varying confounders, and how censoring and missing data were handled (section S.2.1).

Each of the five methods of analysis were applied to estimate the effects of different treatment strategies on $FEV_1$%. Normal-based confidence intervals were obtained based on bootstrap standard errors. Here, we present the results comparing treatment strategies: continue DNase and add hypertonic saline vs continue DNase alone.

### 5.2 Results

We found 4498 individuals who met the eligibility criteria and Figure S.12 in the supplementary file describes how these individuals were selected. Table S.8 in the supplementary file summarises the characteristics of the study population, by treatment combination at time 0. Figure 4 shows the estimated treatment effects, 95% confidence intervals for the effect estimates, and estimated standard errors, for follow-up years 1-5, obtained using each of the five methods. Overall, we found no evidence that adding hypertonic saline has any effect on $FEV_1$%, for people already established on DNase. Across all time points and all methods, almost all 95% confidence intervals contained 0.

Treatment effect and standard error estimates were similar across all methods. As expected, the censoring and weighting approach obtained the largest standard errors, followed by the sequential trials approach. The relatively large standard errors obtained in these methods are due to censoring individuals who deviate from their original treatment strategy. Standard errors obtained by G-estimation were marginally smaller than those obtained via IPTW; G-formula was most efficient. G-formula was the only approach that did not require additional censoring weights (see supplementary file, section S.2.1) which likely explains the smaller standard errors observed in this method.

## 6. Discussion

The aim of this study was to outline how methods that are used to deal with time-dependent confounding can be applied when there are multiple treatments, and to compare them in a simulation study. We described how to implement five different methods for handling time-dependent confounding when estimating the effects of multiple treatments and compared the performance of methods in a simulation study under several scenarios. We also provided an example of implementation using data from the UK CF Registry.

In most of our simulated scenarios, all methods tended to obtain unbiased effect estimates, except for one scenario in which the g-methods lead to biased estimates in the scenario. In this scenario, with treatment effects decreasing over time, the models used in the g-methods were misspecified as our outcome models assumed constant effects over time. To allow for changing effects of treatment over time when using IPTW-MSM and g-formula, an interaction term between treatment and time could be included in the outcome model. For g-estimation, the SNMM can be specified in such a way that decreasing effects are possible (see equation (9)). Censoring and weighting, and the sequential trials approach, were able to obtain unbiased estimates of the decreasing effects without modification to the MSM. This highlights the need to consider what is being assumed in the models required under each approach, and to consider different model specifications.

In terms of efficiency, our results suggested that censoring and weighting, and the sequential trials approach, tended to be the least efficient, while IPTW and g-estimation were the most efficient. This apparently contradicts previous work that suggests IPTW is the least efficient of the g-methods [27,52]. However, our scenarios were relatively simple with one time-varying confounder. When we increased the strength of confounding (scenario 4), IPTW was the less efficient than the g-formula and g-estimation. This is in keeping with previous literature that suggests the inefficiency of IPTW increases as the number of confounders increase [27].

Our applied example investigated the effects of multiple treatment strategies involving different mucoactive nebulisers in people with cystic fibrosis (CF). Questions involving multiple treatments is particularly relevant in the field of CF, since people with CF experience a high treatment burden, and findings ways to reduce treatment burden is a top research priority [60]. This could be achieved by investigating which treatments can be withdrawn, without negatively impacting a person's health, a question which has been brought to prominence in CF over recent years with the widespread introduction of CFTR modulator treatments into routine care, although our data pre-date this era. This type of question has potential applications across all areas of medicine. Polypharmacy (i.e., prescription of multiple treatments) is common, particularly among the older population, and is associated with negative health outcomes and high health care burden [61]. Questions involving multiple treatments may be useful in findings ways to reduce excessive polypharmacy.

A limitation of our study is the results of simulation studies can depend on the way the data are generated. We have tried to overcome this by considering a range of different data generating models. Our scenarios are relatively simple and not necessarily reflective of real data. Future work could consider more complicated scenarios with a higher number of time-varying confounders. Additionally, we did not include the full treatment and covariate history in the propensity score, or outcome models in our analysis of the simulated data. Therefore, our models were conservative, and we might expect larger standard errors, if entire histories are included in the models. In practice, decisions about model specification will involve a trade-off between model complexity and variance in resulting estimates. In the real-world data example, we used last observation carried forward to handle missing data. An advantage of this approach is that it is simple to implement and allowed a fair comparison of results across different methods applied to a complete dataset. However, in general last observation carried forward is not recommended as it is susceptible to bias. Alternative approaches to handling missing data in an analysis using G-methods have been studied recently, including use of weights [40] or multiple imputation [62]. Finally, not all possible methods were considered in this study. We focussed on methods that are commonly used to handle time-dependent confounding in the applied literature [54,63], and considered how to extend these to settings with multiple treatments. Future work could consider the comparison of the methods presented here, to more recent methods which have been proposed for comparing multiple treatment effects using longitudinal data [36,39,40].

In conclusion, several methods exist for estimating the effects of multiple treatment strategies in situations with time-dependent confounding. Such methods may be particularly useful for addressing questions in populations who are taking many treatments, including understanding if there may be scope to rationalise treatments without adversely affecting health outcomes. There are strengths and limitations associated with each of the methods, and the best method for analysis will likely depend on the data and research question.

## Acknowledgements

The authors would like to thank the Cystic Fibrosis Trust for the use of patient registry data to conduct the applied example in this study. The authors also thank the patients, care providers and clinic coordinators at CF centres throughout the UK for their contributions to the patient registries.

## Data Availability

Data may be obtained from a third party and are not publicly available. To access the data, an application must be made to the UK CF Registry Research Committee. [https://www.cysticfibrosis.org.uk/the-work-we-do/uk-cf-registry/apply-for-data-from-the-uk-cf-registry](https://www.cysticfibrosis.org.uk/the-work-we-do/uk-cf-registry/apply-for-data-from-the-uk-cf-registry).

## Funding

GD is supported by a UKRI Future Leaders fellowship (MR/T041285/1). RHK and EG are supported by a UKRI Future Leaders fellowship (MR/S017968/1) awarded to RHK. The funders had no role in the writing, decision to publish, or any other aspect of this manuscript.

## Competing Interest

GD reports speaker honoraria from Chiesi Ltd and Vertex Pharmaceuticals, and advisory board and clinical trial leadership roles with Vertex Pharmaceuticals. RHK reports a speaker honorarium from Vertex Pharmaceuticals.

## Supporting Information

Figures 1-4 referenced in Section 4.1, 4.2 and 5.2 are located here. Supplementary information is located later in this document. R code for the simulation study can be found at: https://github.com/EmilyG602/MultipleTreatmentsSimulationStudy

*Figure 1: Directed Acyclic Graph (DAG) depicting causal relationships between variables in the simulated data*

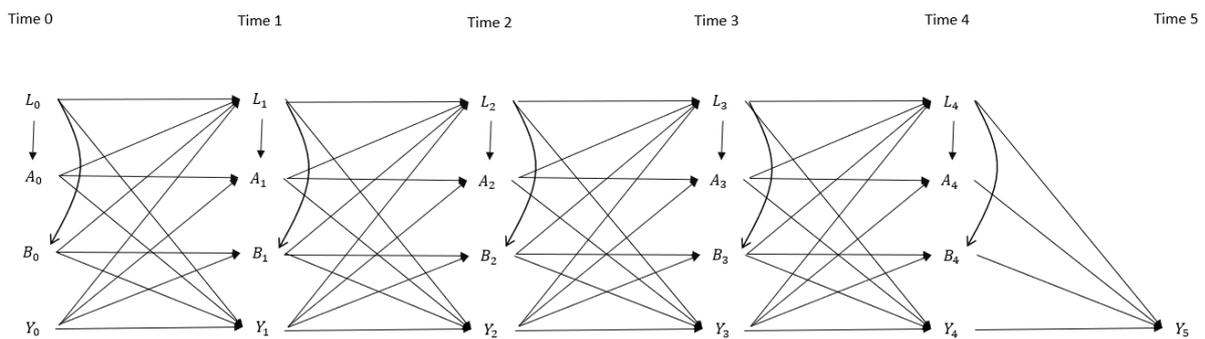

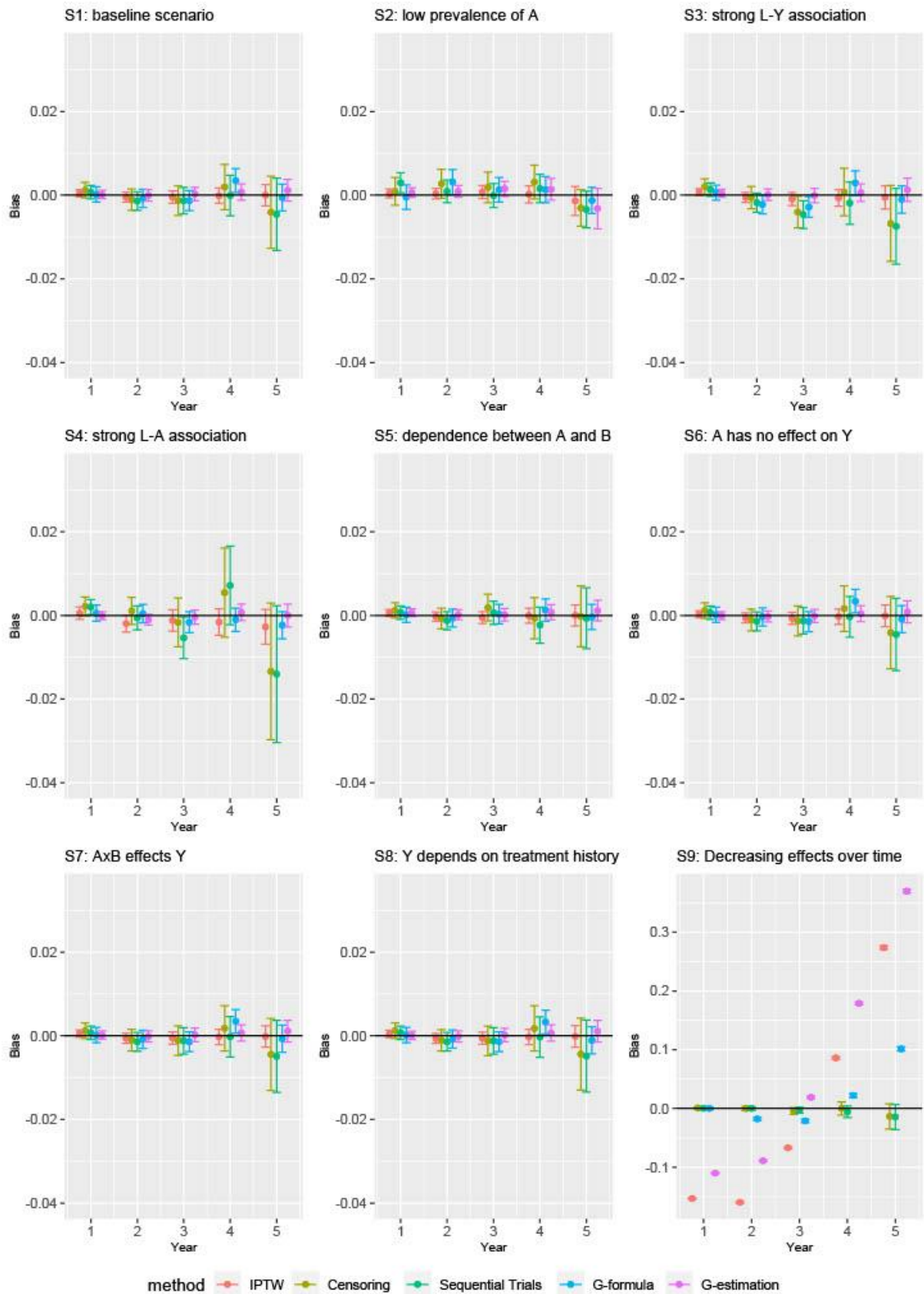

Figure 2: Estimated bias (and 95% confidence intervals) obtained when estimating the effect of treatments A and B, compared to B only, in simulated scenarios 1-9.

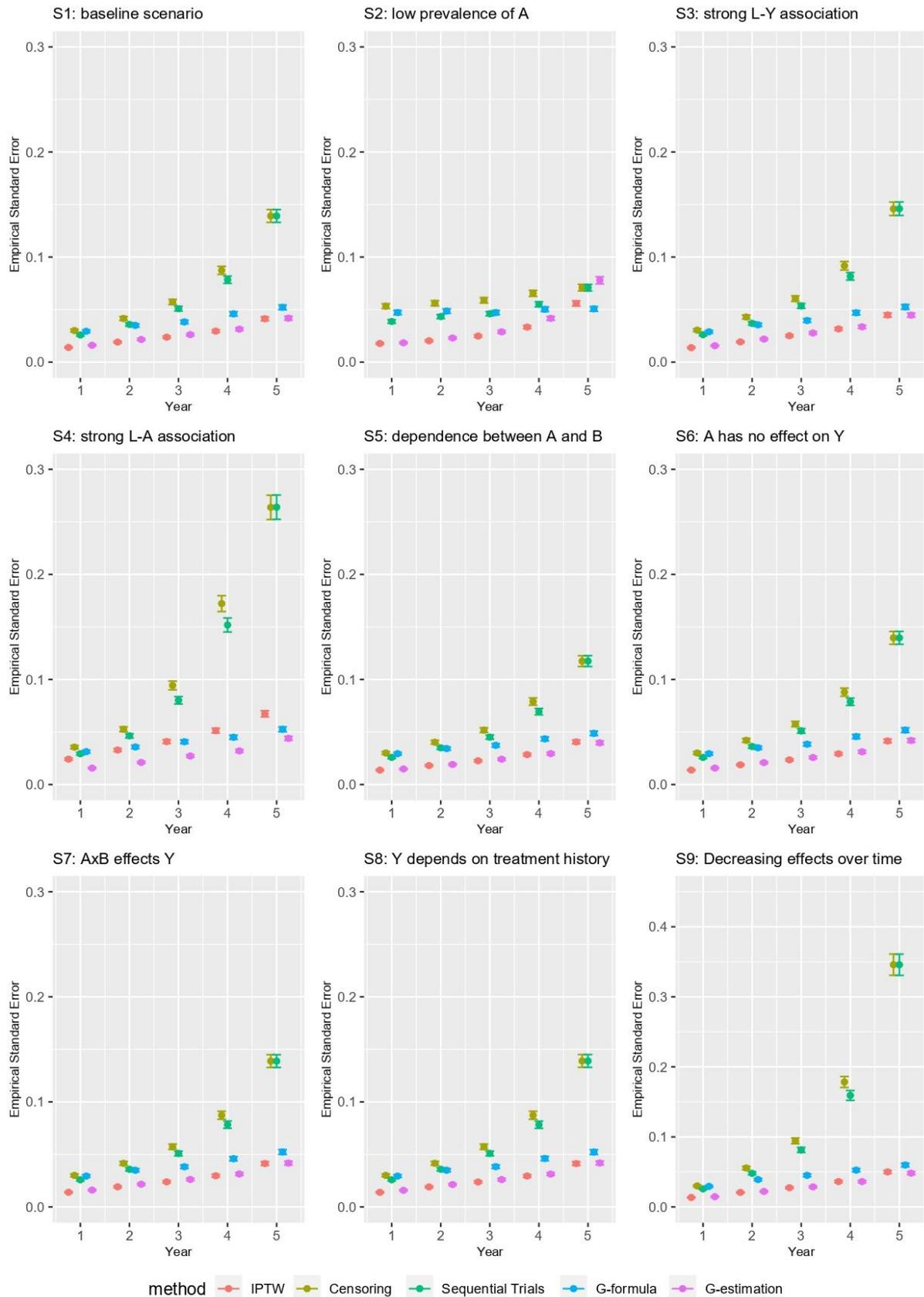

*Figure 3: Estimated standard error (and 95% confidence intervals) obtained when estimating the effect of treatments A and B, compared to B only, in simulated scenarios 1-9.*

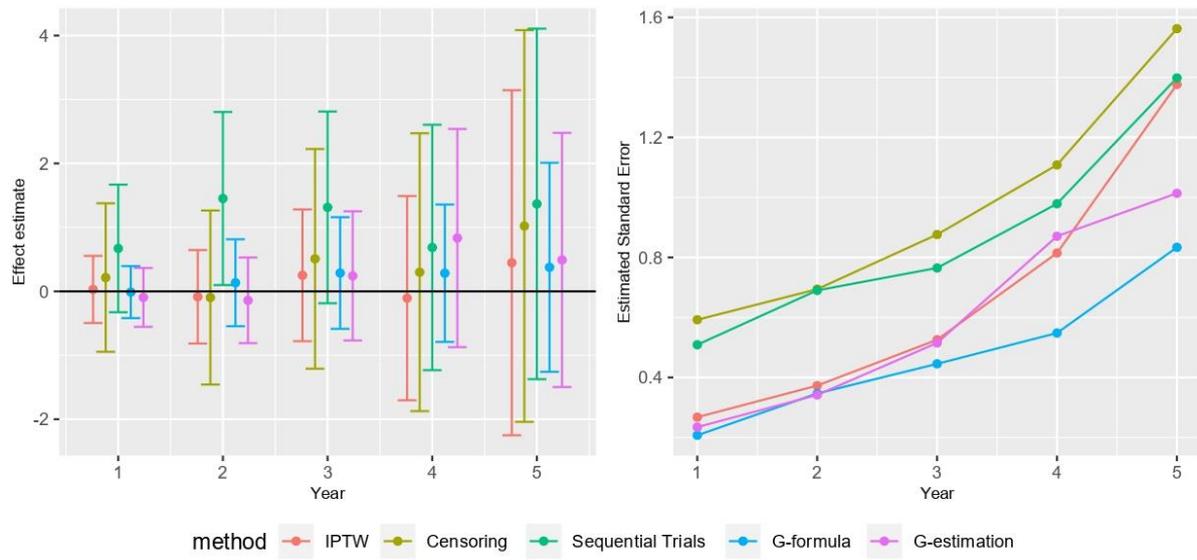

Figure 4: Estimated treatment effects of taking DNase and hypertonic saline, compared to DNase alone, for years 1-5, and their associated standard errors.

# SUPPLEMENTARY MATERIAL

# Investigating the causal effects of multiple treatments using longitudinal data: a simulation study

This document consists of two sections: (1) additional notes on the methods used and results obtained in the simulation study presented in the main manuscript and (2) additional notes on the methods used and results obtained in the real data application. The current document is organised as follows:

*S.1 Additional notes on the simulation study*
 *S.1.1 Data Generation*
 *S.1.2 Further details on the implementation of methods*
 *S.1.3 True treatment effects*
 *S.1.4 Results obtained for alternative comparisons*

*S.2 Additional notes on the illustrative example using data from the UK CF Registry*
 *S.2.1 Further details on the implementation of methods*
 *S.2.2 Additional results*

## S.1 Additional notes on methodology
### S.1.1 Data Generation

All data generation and analyses were conducted using R Version 3.6.1. Data were simulated under nine scenarios. For each scenario, 1000 datasets were iteratively simulated. The starting seed number was the same in each scenario (06122020).

Datasets were simulated with 10,000 individuals and 6 time points ($t = 0, ...,5$). Four time-varying variables were simulated. This included two binary treatment variables, $A$, and $B$, a continuous outcome, $Y$, and a continuous confounder, $L$.

Baseline confounders measured at time 0 were simulated using the following formulae:

$$L_0 \sim N(0,1); \quad A_0 = 0; \quad B_0 = 0; \quad Y_0 \sim N(0.05L_0, 1) \tag{S.1}$$

Treatment at baseline (i.e., time 0) was simulated as follows (see Table S.1 for values of the coefficients):

$$A_0 \sim Bin(1, expit(\beta_0 + \beta_1 L_0)) \tag{S.2}$$
$$B_0 \sim Bin(1, expit(\gamma_0 + \gamma_1 L_0)) \tag{S.3}$$

For Scenarios 1-5, data at times 0-4 were simulated sequentially with the following formulae, where the regression coefficients were varied between scenarios (see Table S.1):

$$L_t \sim N(0.2L_{t-1} + 0.2A_{t-1} + 0.2B_{t-1} + 0.01Y_{t-1}, 1) \tag{S.4}$$

$$A_t \sim Bin(1, expit(\beta_0 + \beta_1 L_t + \beta_2 A_{t-1} + \beta_3 B_{t-1} + \beta_4 Y_{t-1})) \tag{S.5}$$

$$B_t \sim Bin(1, expit(\gamma_0 + \gamma_1 L_t + \gamma_2 A_{t-1} + \gamma_3 B_{t-1} + \gamma_4 Y_{t-1})) \tag{S.6}$$

$$Y_{t+1} \sim N(\delta_0 + \delta_1 L_t + \delta_2 A_t + \delta_3 B_t + \delta_4 Y_t, 1) \tag{S.7}$$

*Table S.1: Coefficients used in equations (S.1)-(S.4) in the generation of data for scenarios 1-5*

| Model | Intercept | Coefficient for: | | | |
|---|---|---|---|---|---|
| | | L | A | B | Y |
| Scenario 1 (baseline scenario) | | | | | |
| $A_t$ | 0 | 0.30 | 1.80 | 0 | 0.12 |
| $B_t$ | 0 | 0.30 | 0 | 1.80 | 0.12 |
| $Y_t$ | 0 | 0.05 | 1.00 | 0.50 | 0.10 |
| Scenario 2 (low prevalence of $A$) | | | | | |
| $A_t$ | -2.3 | 0.30 | 1.80 | 0 | 0.12 |
| $B_t$ | 0 | 0.30 | 0 | 1.80 | 0.12 |
| $Y_t$ | 0 | 0.05 | 1.00 | 0.50 | 0.10 |
| Scenario 3 (strong L-Y association) | | | | | |
| $A_t$ | 0 | 0.30 | 1.80 | 0 | 0.12 |
| $B_t$ | 0 | 0.30 | 0 | 1.80 | 0.12 |
| $Y_t$ | 0 | 0.30 | 1.00 | 0.50 | 0.10 |
| Scenario 4 (strong L-A association) | | | | | |
| $A_t$ | 0 | 1.00 | 1.80 | 0 | 0.12 |
| $B_t$ | 0 | 0.30 | 0 | 1.80 | 0.12 |
| $Y_t$ | 0 | 0.05 | 1.00 | 0.50 | 0.10 |
| Scenario 5 (dependence between A and B) | | | | | |
| $A_t$ | 0 | 0.30 | 1.80 | -0.2 | 0.12 |
| $B_t$ | 0 | 0.30 | -0.2 | 1.80 | 0.12 |
| $Y_t$ | 0 | 0.05 | 1.00 | 0.50 | 0.10 |

For scenarios 6-9, data were again simulated sequentially, and the values for $L$, $A$, and $B$ were simulated using the same model parameters as in scenario 1. However, the specification of the outcome models varied:

Scenario 6 ($A$ has no effect on $Y$):

$$Y_{t+1} \sim N(0.05L_t + 0.5B_t + 0.1Y_t, 1) \tag{S.8}$$

Scenario 7 (the interaction between $A$ and $B$ has an effect on $Y$):

$$Y_{t+1} \sim N(0.05L_t + A_t + 0.5B_t + 0.25A_t B_t + 0.01Y_t, 1) \tag{S.9}$$

Scenario 8 ($Y_t$ depends on treatment history):

$$Y_{t+1} \sim N(0.05L_t + A_t + 0.2A_{t-1} + 0.1A_{t-2} + 0.05A_{t-3} + 0.001A_{t-4} + 0.5B_t + 0.1Y_t, 1) \tag{S.10}$$

Scenario 9 (decreasing effects over time):

$$Y_{t+1} \sim N(0.05L_t + (1 - 0.2\sum_{j=1}^{t} A_j)A_t + 0.5B_t + 0.1Y_t, 1) \tag{S.11}$$

### S.1.2 Further details on the implementation of methods

Section 4.1 of the accompanying paper describes how the methods were implemented in the simulation study. Here we provide the formulas for the weights and conditional models that are referred to in Section 4.1.

*Inverse-probability-of-treatment weighting*

We used stabilised weights when applying IPTW, defined as follows:

$$weight_t = \prod_{j=1}^{t} \frac{\Pr(Z_j = z_j | Z_{j-1} = z_{j-1}, Y_0 = y_0)}{\Pr(Z_j = z_j | Z_{j-1} = z_{j-1}, Y_0 = y_0, Y_{j-1} = y_{j-1}, L_{j-1} = l_{j-1})} \tag{S.12}$$

The probabilities required for the weights were obtained using logistic regression.

*Censoring and weighting and the sequential trials approach*

The weights defined in S.12 were also used as censoring weights in the censoring and weighting method to account for censoring due to treatment switching. For the sequential trials approach, the weights were similar, except an indicator for trial was including in the logistic models predicting treatment.

*The g-formula*

Implementation of the g-formula requires postulating models for the time-varying confounders as well as the outcome and treatment. The conditional models for time-varying confounders, $L_t$ and $Y_t$, were defined as follows:

$$L_t = \gamma_0 + \gamma_L L_{t-1} + \sum_{c=1}^{3} \gamma_{cz} I(z_{t-1} = c) + \gamma_Y Y_{t-1} + \varepsilon_t \tag{S.13}$$

$$Y_t = \tau_0 + \tau_L L_{t-1} + \sum_{j=1}^{t-1}\sum_{c=1}^{3} \tau_{cj} I(z_j = c) + \tau_Y Y_{t-1} + \varepsilon_t \tag{S.14}$$

### S.1.3 True treatment effects

Tables S.2-S.7 collectively provide the true treatment effects for all treatment combination comparisons, all times points, and all scenarios.

*Table S.2: True treatment effects for all estimands at all time points in scenarios 1, 2, 4, 5*

| Comparison | Year 1 | Year 2 | Year 3 | Year 4 | Year 5 |
|---|---|---|---|---|---|
| A vs Nil | 1.00 | 1.11 | 1.12 | 1.13 | 1.13 |
| B vs Nil | 0.50 | 0.56 | 0.57 | 0.57 | 0.57 |
| A vs B | -0.50 | -0.55 | -0.55 | -0.56 | -0.56 |
| A & B vs Nil | 1.50 | 1.67 | 1.69 | 1.70 | 1.70 |
| A & B vs A | 0.50 | 0.56 | 0.57 | 0.57 | 0.57 |
| A & B vs B | 1.0 | 1.11 | 1.12 | 1.13 | 1.13 |

*Nil denotes no treatment*

*Table S.3: True treatment effects for all estimands at all time points in scenario 3*

| Comparison | Year 1 | Year 2 | Year 3 | Year 4 | Year 5 |
|---|---|---|---|---|---|
| A vs Nil | 1.00 | 1.16 | 1.19 | 1.20 | 1.20 |
| B vs Nil | 0.50 | 0.61 | 0.64 | 0.64 | 0.64 |
| A vs B | -0.50 | -0.55 | -0.56 | -0.56 | -0.56 |
| A & B vs Nil | 1.50 | 1.77 | 1.83 | 1.84 | 1.84 |
| A & B vs A | 0.50 | 0.61 | 0.64 | 0.64 | 0.64 |
| A & B vs B | 1.00 | 1.16 | 1.19 | 1.20 | 1.20 |

*Nil denotes no treatment*

*Table S.4: True treatment effects for all estimands at all time points in scenario 6*

| Comparison | Year 1 | Year 2 | Year 3 | Year 4 | Year 5 |
|---|---|---|---|---|---|
| A vs Nil | 0.00 | 0.01 | 0.01 | 0.01 | 0.01 |
| B vs Nil | 0.50 | 0.56 | 0.57 | 0.57 | 0.57 |
| A vs B | 0.50 | 0.55 | 0.56 | 0.56 | 0.56 |
| A & B vs Nil | 0.50 | 0.57 | 0.58 | 0.58 | 0.58 |
| A & B vs A | 0.50 | 0.56 | 0.57 | 0.57 | 0.57 |
| A & B vs B | 0.00 | 0.01 | 0.01 | 0.01 | 0.01 |

*Nil denotes no treatment*

*Table S.5: True treatment effects for all estimands at all time points in scenario 7*

| Comparison | Year 1 | Year 2 | Year 3 | Year 4 | Year 5 |
|---|---|---|---|---|---|
| A vs Nil | 1.00 | 1.11 | 1.12 | 1.13 | 1.13 |
| B vs Nil | 0.50 | 0.56 | 0.57 | 0.57 | 0.57 |
| A vs B | -0.50 | -0.55 | -0.56 | -0.56 | -0.56 |
| A & B vs Nil | 1.75 | 1.95 | 1.97 | 1.97 | 1.97 |
| A & B vs A | 0.75 | 0.84 | 0.85 | 0.85 | 0.85 |
| A & B vs B | 1.25 | 1.39 | 1.40 | 1.40 | 1.40 |

*Nil denotes no treatment*

*Table S.6: True treatment effects for all estimands at all time points in scenario 8*

| Comparison | Year 1 | Year 2 | Year 3 | Year 4 | Year 5 |
|---|---|---|---|---|---|
| A vs Nil | 1.00 | 1.31 | 1.44 | 1.51 | 1.52 |
| B vs Nil | 0.50 | 0.56 | 0.57 | 0.57 | 0.57 |
| A vs B | -0.50 | -0.75 | -0.88 | -0.94 | -0.95 |
| A & B vs Nil | 1.50 | 1.87 | 2.01 | 2.08 | 2.08 |
| A & B vs A | 0.50 | 0.56 | 0.57 | 0.57 | 0.57 |
| A & B vs B | 1.00 | 1.31 | 1.44 | 1.51 | 1.52 |

*Nil denotes no treatment*

*Table S.7: True treatment effects for all estimands at all time points in scenario 9*

| Comparison | Year 1 | Year 2 | Year 3 | Year 4 | Year 5 |
|---|---|---|---|---|---|
| A vs Nil | 1.00 | 0.91 | 0.70 | 0.48 | 0.26 |
| B vs Nil | 0.50 | 0.56 | 0.57 | 0.57 | 0.57 |
| A vs B | -0.50 | -0.35 | -0.14 | 0.09 | 0.31 |
| A & B vs Nil | 1.50 | 1.47 | 1.27 | 1.05 | 0.83 |
| A & B vs A | 0.50 | 0.56 | 0.57 | 0.57 | 0.57 |
| A & B vs B | 1.00 | 0.91 | 0.70 | 0.48 | 0.26 |

*Nil denotes no treatment*

## S.1.4 Results obtained for alternative comparisons

The relative trends in performance between the five methods, in terms of bias and standard error, was similar in all treatment combination comparisons (see Figures S.1-S.10).

Regarding bias, all methods tended to obtain unbiased estimates in scenarios 1-8 whereas in scenario 9, bias was observed in the estimates obtained using IPTW, g-formula, or g-estimation. One notable exception was for the comparison treatment $A$ only to no treatment (i.e. $ATE_t^{\{A-0\}}$), in scenario 3. In this case, IPTW obtained small, but consistent bias for all time points. In scenario 3, there was a stronger association between $L$ and $Y$. It is likely that the weights used in IPTW were unable to remove all imbalances in the distribution of $L$ between groups of individuals taking $A$ and not taking $A$, and due to the strong $L-Y$ association, and the remaining imbalances will have led to residual confounding bias.

Regarding standard error, the following trends were observed in most treatment comparisons and scenarios: the rate of increase in standard error by time was greatest for sequential trials and censoring and weighting, leading to larger standard errors in these methods by time point 3. The rate of increase in standard error by time was similar in g-formula, IPTW and g-estimation, however the magnitude of standard error was consistently greater in g-formula, compared to IPTW and g-estimation. IPTW and g-estimation performed similarly in all scenarios, except scenario 4, for which g-estimation was most efficient. The only exception to the above trends was for the comparison of treatment $A$ and $B$ to no treatment (i.e. $ATE_t^{\{AB-0\}}$). In this case, the sequential trials method, and censoring and weighting method, tended to be as efficient as IPTW and g-estimation. This is because individuals on no treatment, or treatment combination $A$ and $B$ are less likely to switch treatment combination compared to those taking $A$ only or $B$ only (due to the way the data are simulated). Therefore, when comparing treatment $A$ and $B$ to no treatment, less individuals are censored, leading to greater sample sizes and smaller standard errors.

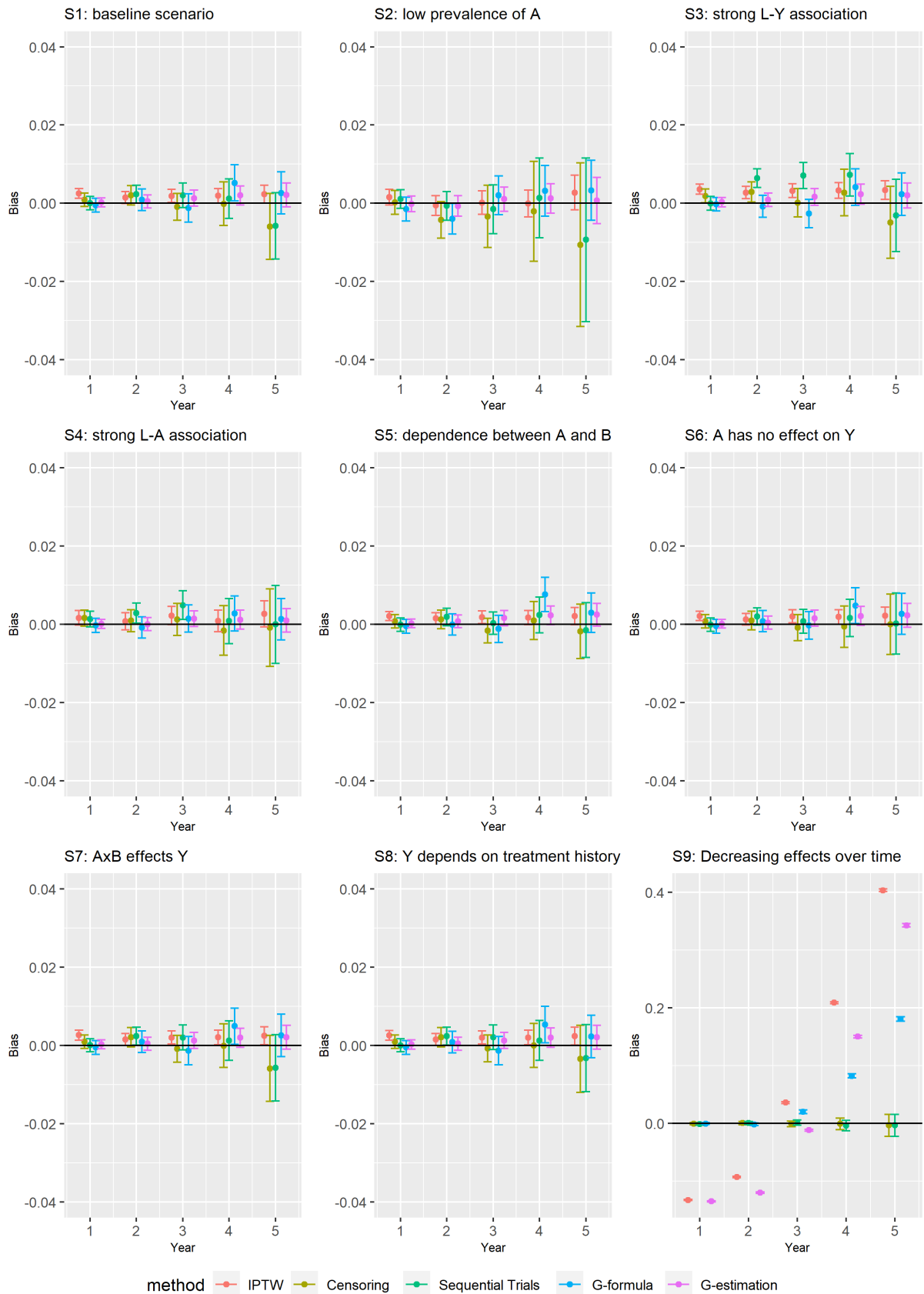

Figure S.2: Estimated bias (and 95% confidence intervals) obtained when estimating the effect of treatment $A$ versus no treatment, in simulated scenarios 1-9.

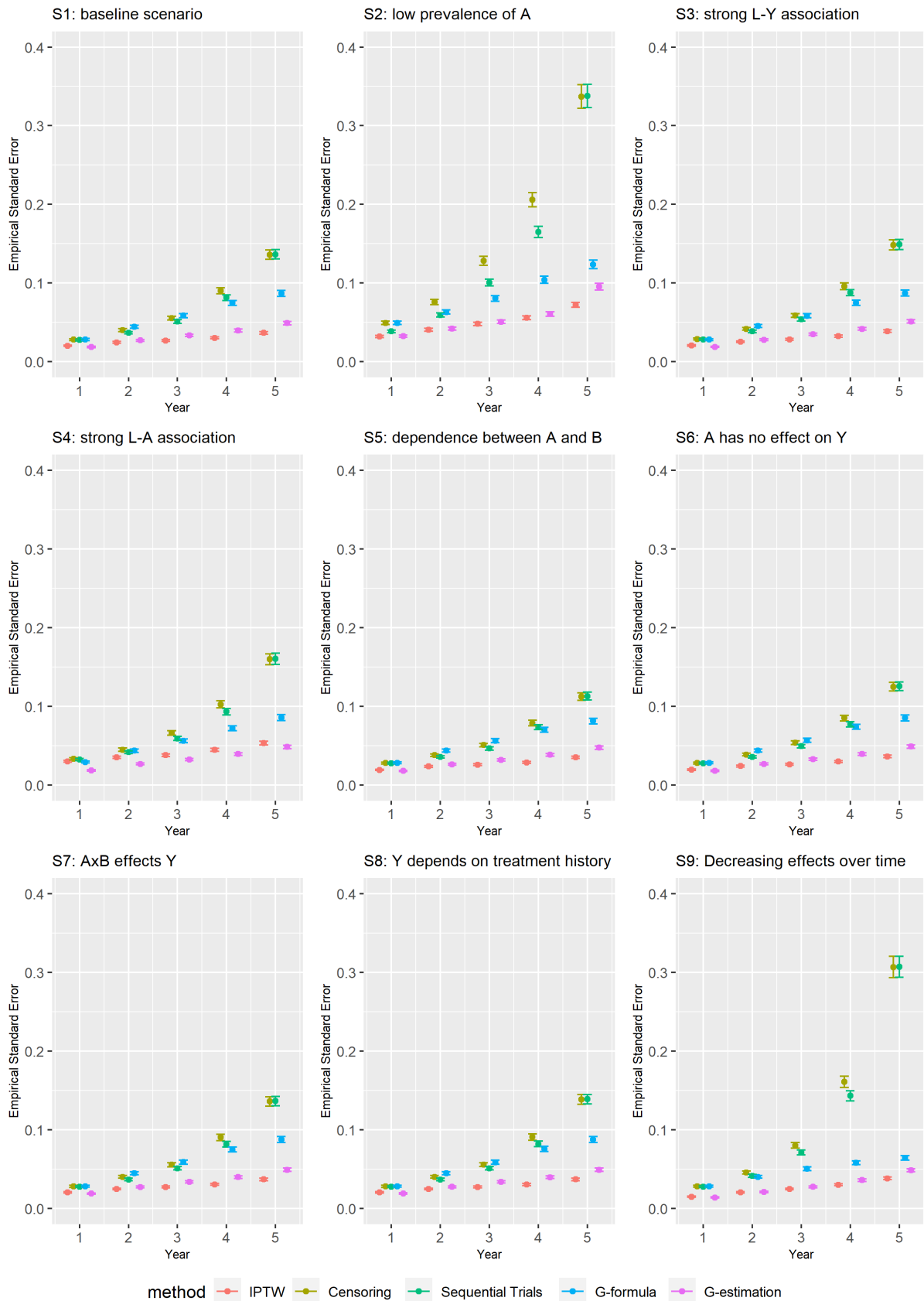

Figure S.3: Estimated standard error (and 95% confidence intervals) obtained when estimating the effect of A versus no treatment in Scenarios 1-9

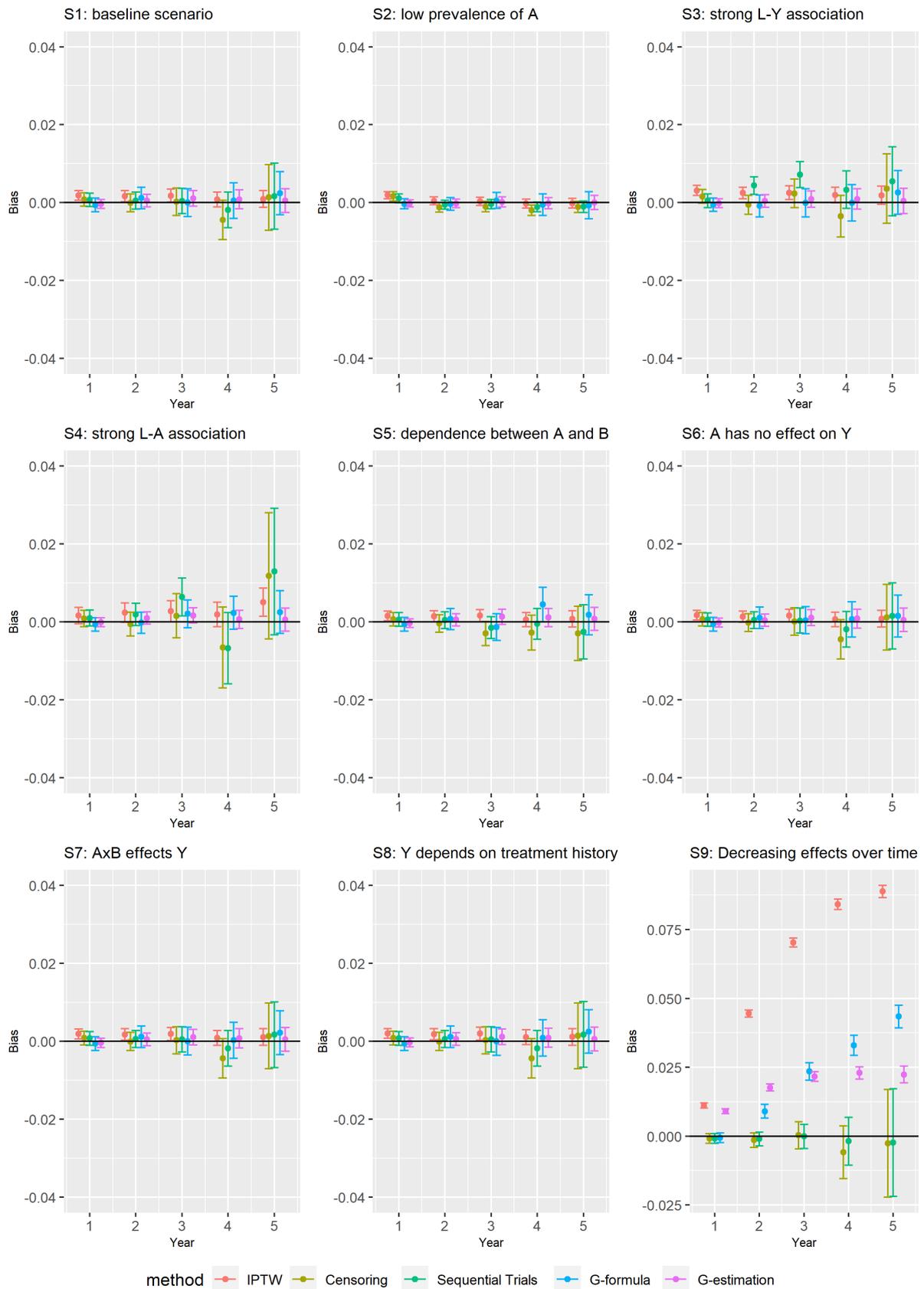

Figure S.3: Estimated bias (and 95% confidence intervals) obtained when estimating the effect of treatment $B$ versus no treatment, in simulated scenarios 1-9.

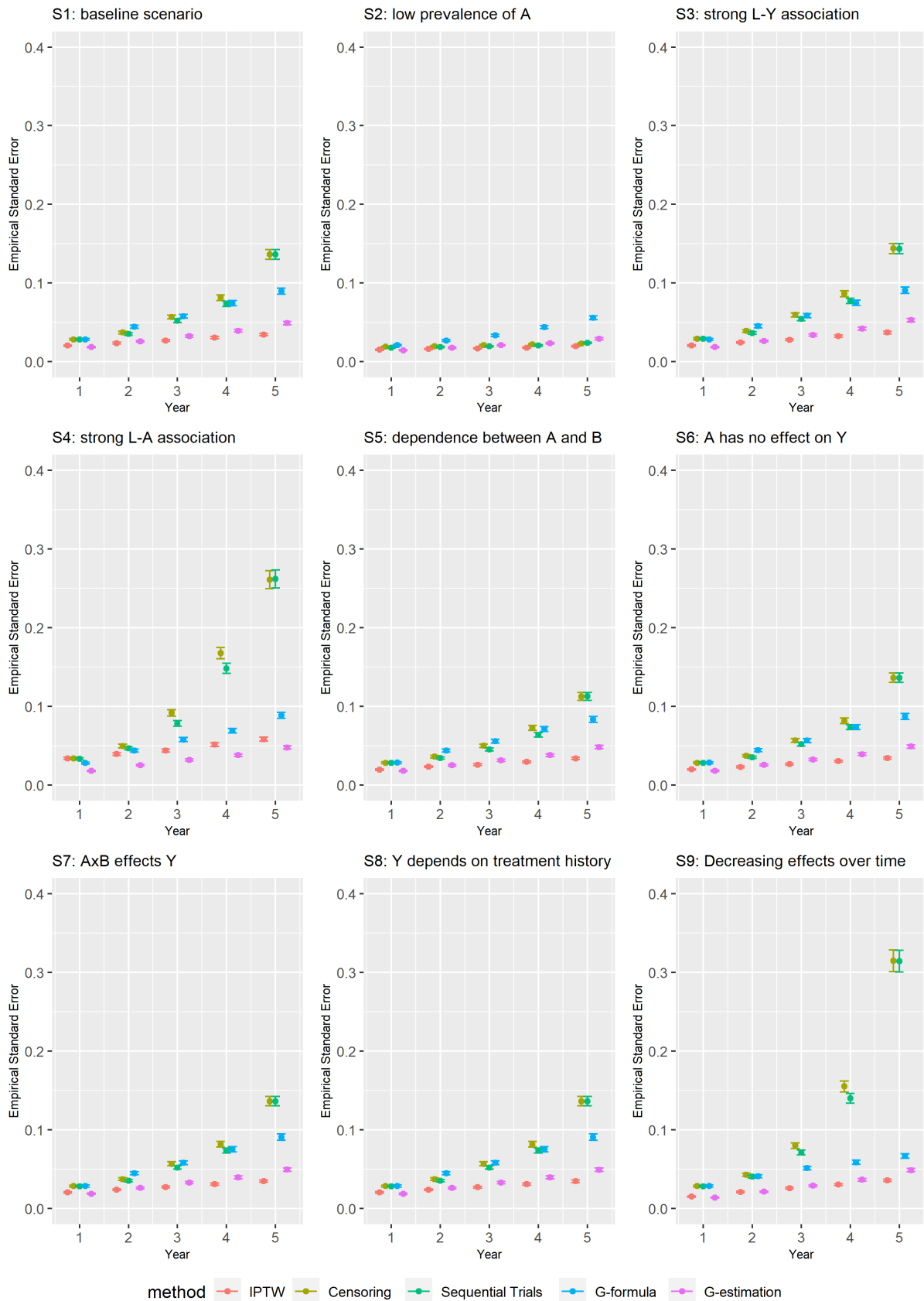

Figure S.4: Estimated standard error (and 95% confidence intervals) obtained when estimating the effect of B versus no treatment in Scenarios 1-9

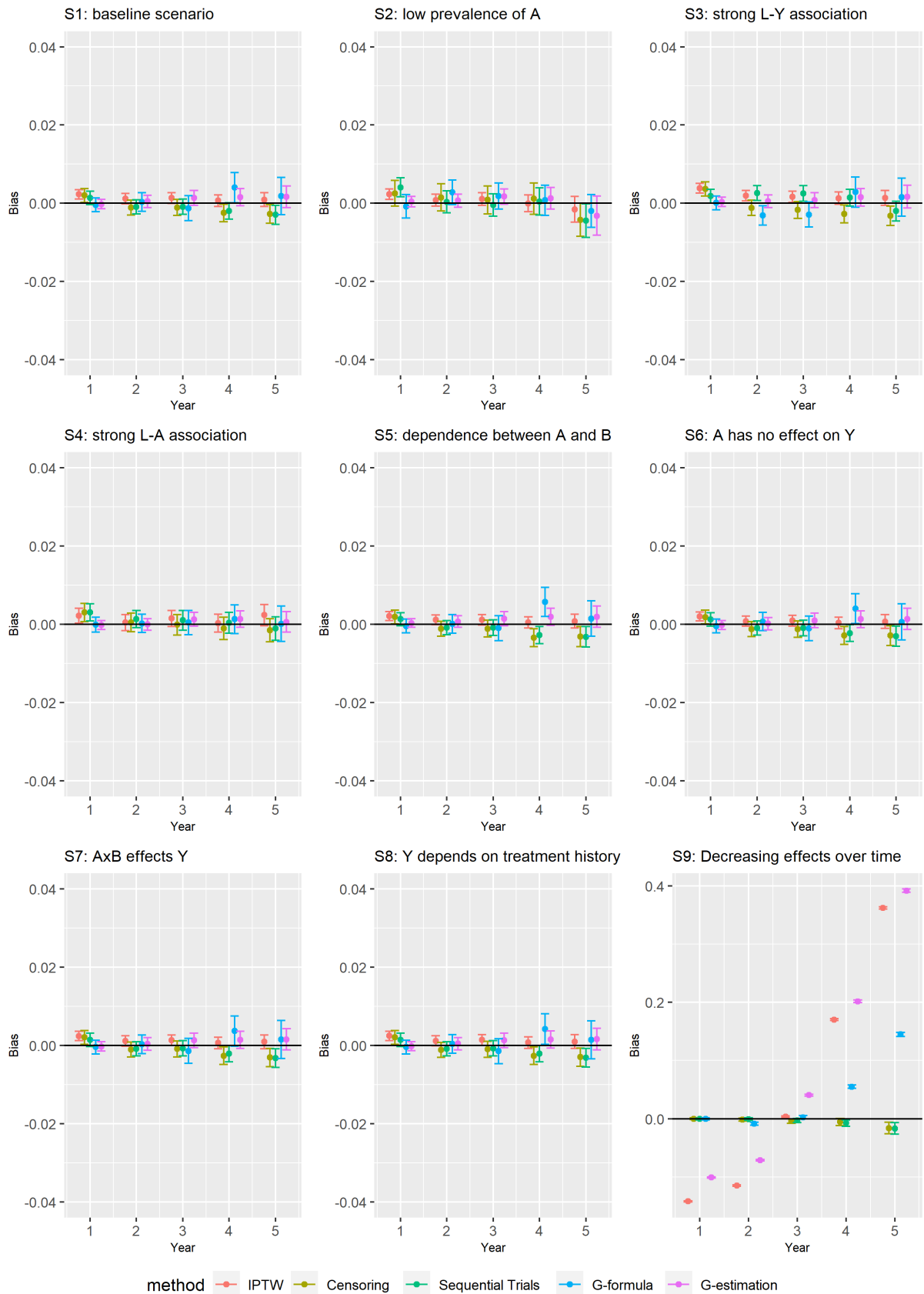

Figure S.5: Estimated bias (and 95% confidence intervals) obtained when estimating the effect of treatments $A$ and $B$ versus no treatment, in simulated scenarios 1-9.

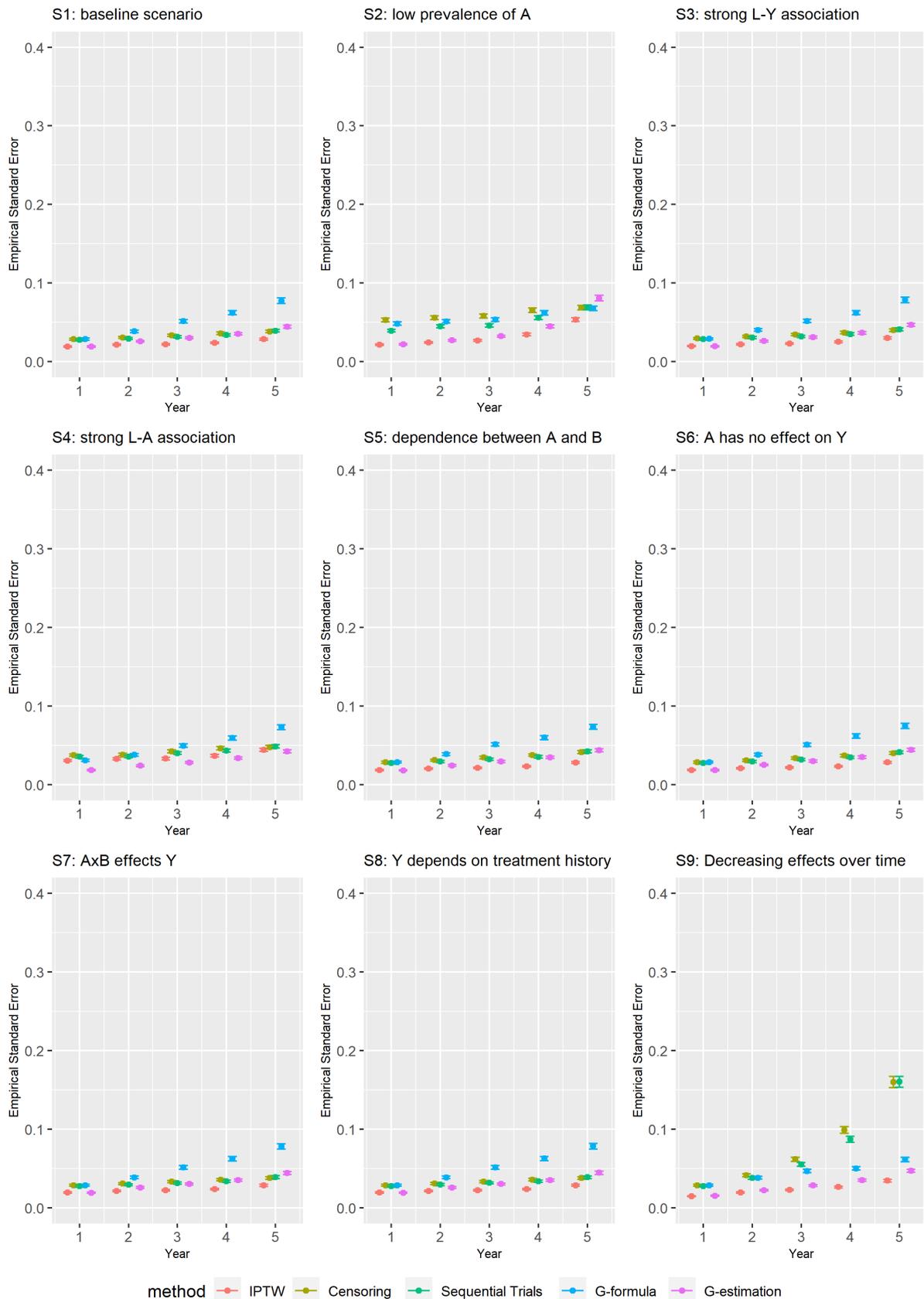

Figure S.6: Estimated standard error (and 95% confidence intervals) obtained when estimating the effect of A and B versus no treatment in Scenarios 1-9

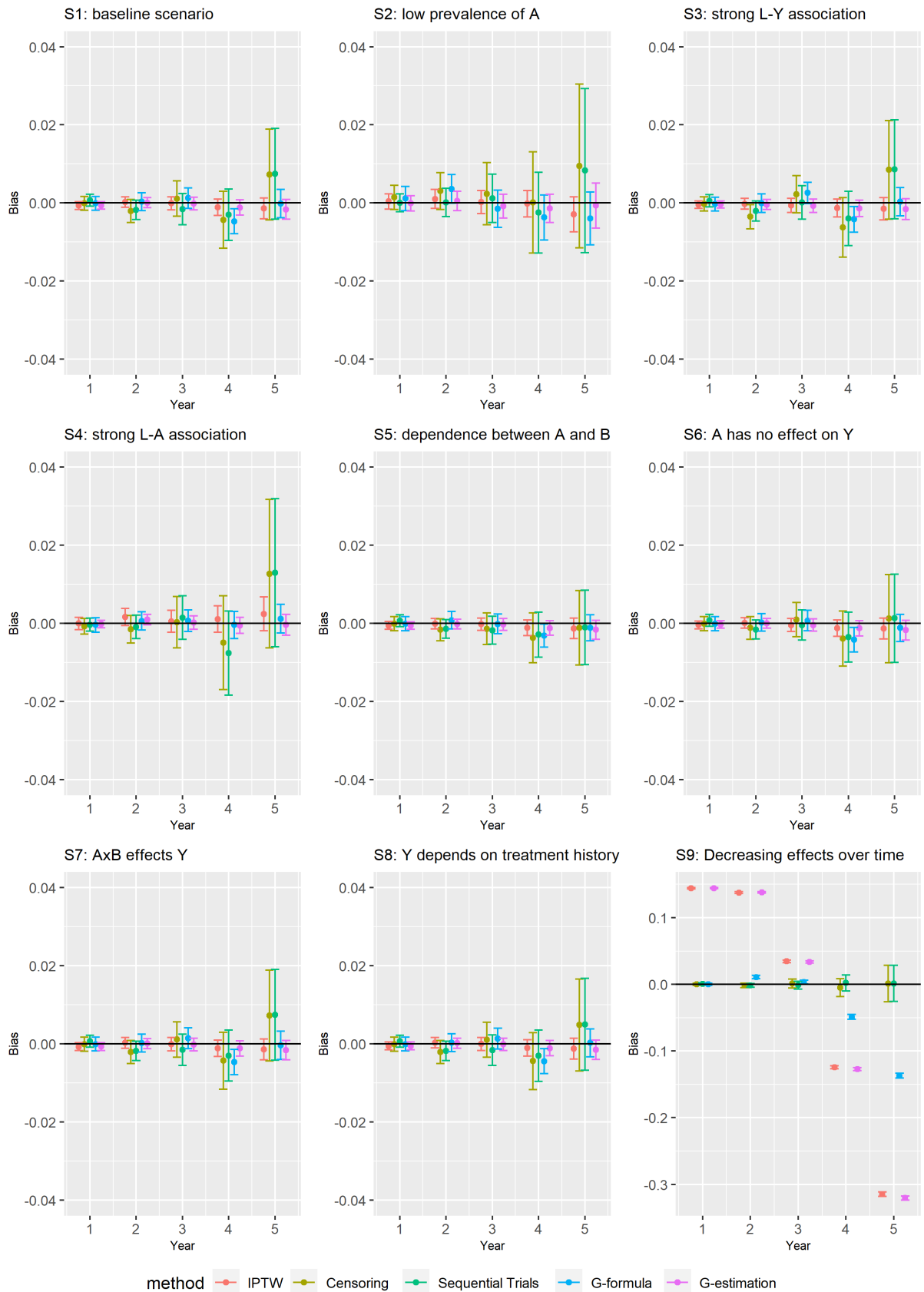

Figure S.7: Estimated bias (and 95% confidence intervals) obtained when estimating the effect of treatment $A$ versus $B$, in simulated scenarios 1-9.

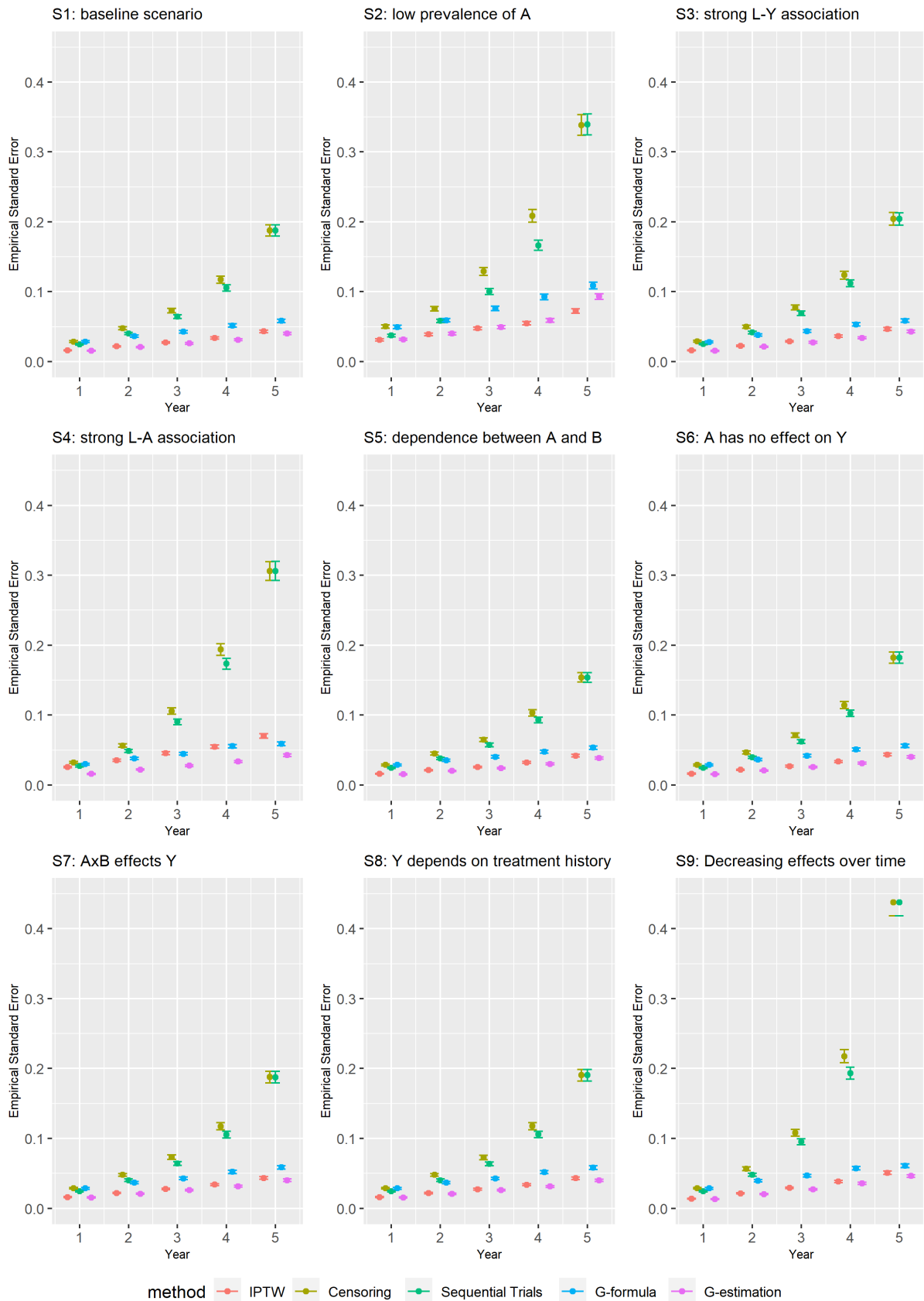

Figure S.8: Estimated standard error (and 95% confidence intervals) obtained when estimating the effect of A versus B in Scenarios 1-9

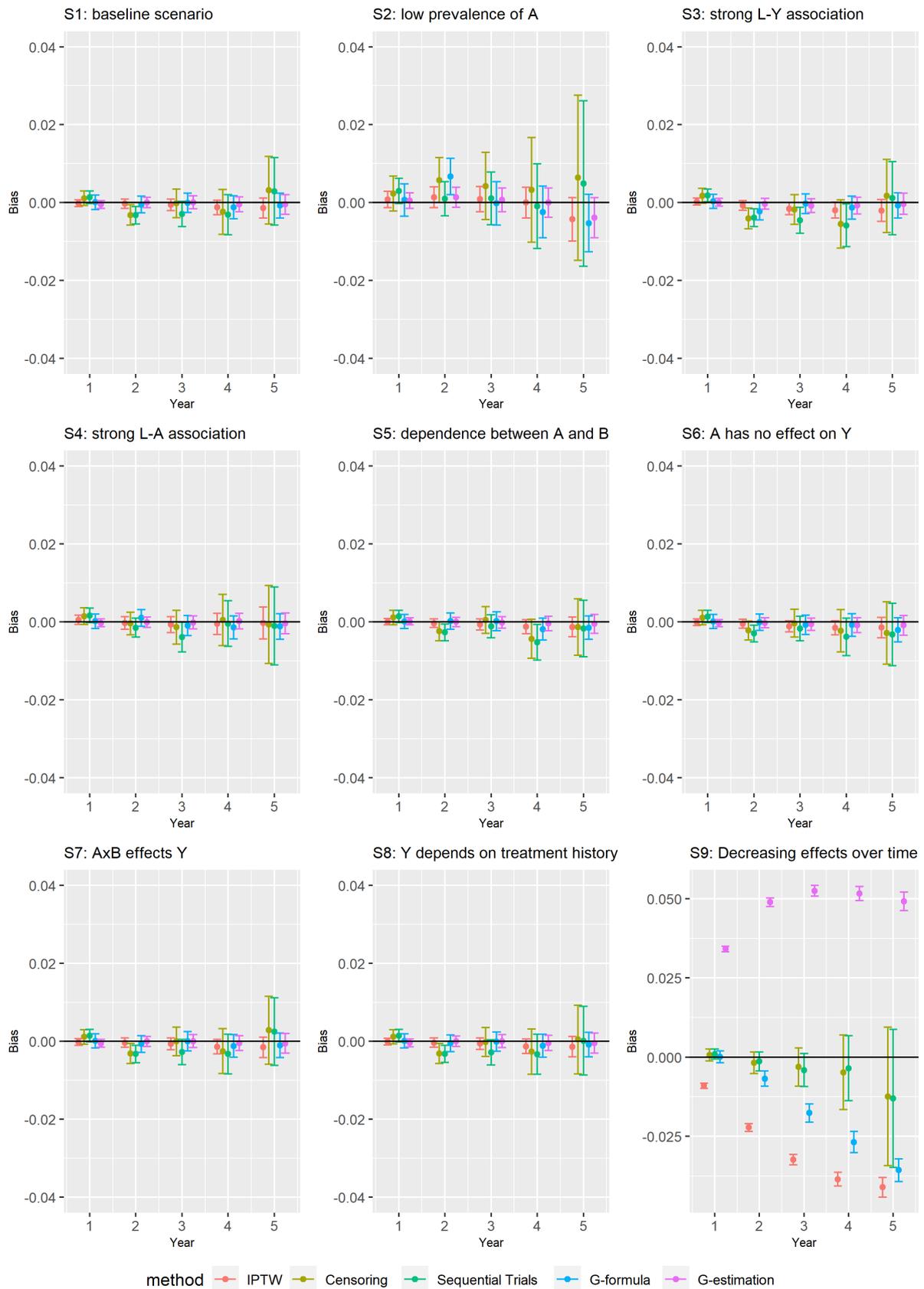

Figure S.9: Estimated bias (and 95% confidence intervals) obtained when estimating the effect of treatments $A$ and $B$ versus $A$ only, in simulated scenarios 1-9.

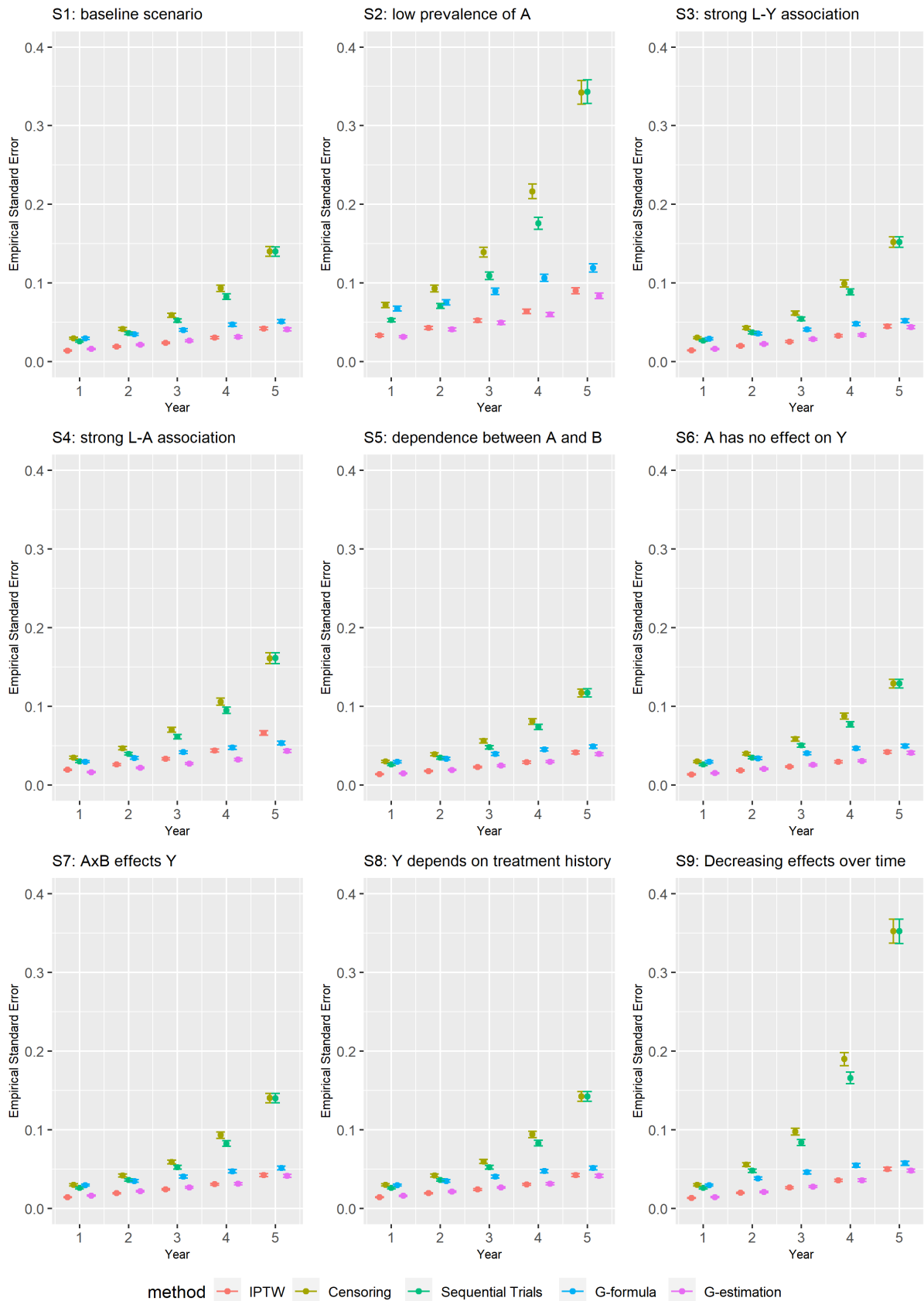

Figure S.10: Estimated standard error (and 95% confidence intervals) obtained when estimating the effect of A and B versus A in Scenarios 1-9

## S.2 Additional notes on the illustrative example using data from the UK CF Registry
### S.2.1 Further details on the implementation of methods
*Confounding variables*

Figure S.11 is the directed acyclic graph (DAG) which shows the assumed relationships between variables in our data for the applied analysis. Time 0 is the time at which the eligibility criteria are met and individuals enter the study. At this time, individuals may follow one of the following treatment strategies:

$$Z_t = \begin{cases} 0 \text{ if } DNase_t = 0 \text{ and } Hypertonic\ saline_t = 0 \\ 1 \text{ if } DNase_t = 1 \text{ and } Hypertonic\ saline_t = 0 \\ 2 \text{ if } DNase_t = 0 \text{ and } Hypertonic\ saline_t = 1 \\ 3 \text{ if } DNase_t = 1 \text{ and } Hypertonic\ saline_t = 1 \end{cases} \quad (S.15)$$

The outcome, FEV$_1$%, at time $t$ is denoted by $Y_t$

*Figure S. 11: Directed Acyclic Graph depicting the assumed short-term confounding paths of the treatment-outcome association*

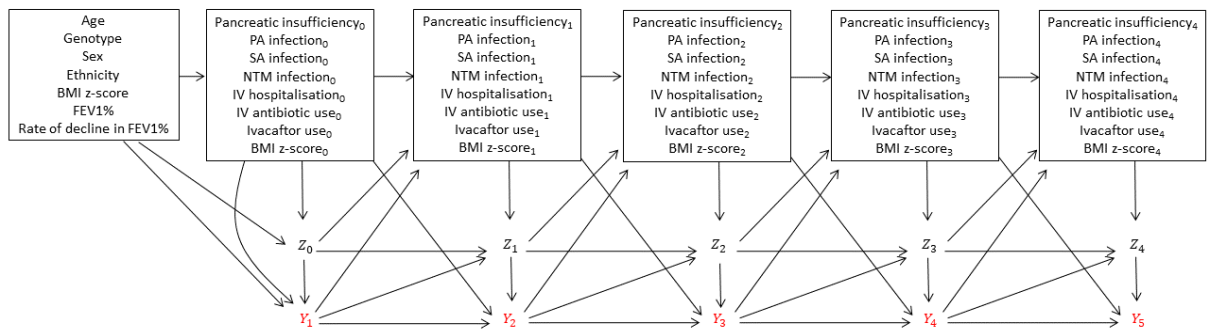

As can be seen from Figure S.11, the variables included as time-invariant confounders were: age at baseline, genotype, sex, ethnicity, BMI z-score at baseline, FEV$_1$% at baseline and rate of decline in FEV$_1$%. The variables included as time-varying confounders were: pancreatic insufficiency, ivacaftor use, *p. Aeruginosa* infection, *staphylococcus aureus* infection, *nontuberculous mycobacteria* infection, hospital admissions for intravenous antibiotics, days on intravenous antibiotics, BMI z-score and past FEV$_1$%.

Genotype was classed as either high risk, low risk or not assigned as previously defined[1]. Ethnicity was classed as white or non-white due to small numbers in non-white ethnic groups in this population. Rate of decline in FEV$_1$% represented the change in FEV$_1$% observed prior to baseline. We defined the following linear mixed model with random slope and intercept:

$$FEV_1\%_{ij} = (\alpha_0 + \delta_{0i}) + (\alpha_1 + \delta_{1i})j + e_{ij} \quad (S.16)$$

Where $j \in \{0,1,2,3,4\}$ is the number of years before baseline ($j = 0$ is the baseline year). The estimate of the slope parameter ($\alpha_1 + \delta_{1i}$) for each individual is used as a time-invariant variable representing rate of change in FEV$_1$%.

Pancreatic insufficiency was a yes/no indicator where individuals were assigned "yes" if they were prescribed pancreatic enzyme supplements. IV hospital admissions was a yes/no indicator were yes indicated individuals had at least one hospital admission for IV antibiotics over the past year. IV days included home and hospital admissions and was categorised as: 0, 1-4, 15-28 and 29+. BMI z-scores were calculated using the WHO reference distribution[2] and FEV$_1$% was calculated using the Global Lung Initiative equations[3].

*Notation*

Let $L_B$ denote the set of time-invariant confounders and $L_t$ denote the set of time-varying confounders recorded at time $t$. Then:

$$L_B = \{Age_0, Genotype, Sex, Ethnicity, Rate\ of\ decline\ in\ FEV1\%, FEV1\%_0 BMI_0\}$$

And $L_t$ is defined as:

$$L_t = \{FEV1\%_{t-1}, BMI_t, IV\ days_t, IV\ Hospital\ Admission_t,$$

$$NTM_t, SA\ infection_t, PA\ infection_t, Ivacaftor\ use_t, Pancreatic\ insufficiency_t\}$$

*Censoring*

Individuals who died, or had an organ transplant were censored at the time of the event. We also censored individuals who began treatment with lumacaftor/ivacaftor, texacaftor/ivacaftor, or mannitol, at the time of treatment initiation. Furthermore, some individuals had less than five years of follow-up due to administrative end of follow-up; these individuals were accounted for in the censoring as described below.

When implementing the g-formula with a censored population, we use the same approach as Taubman et al[4]. This involves simulating outcomes for each individual beyond the times when they were censored. Outcomes are evaluated at the end of follow-up, assuming that it occurs at the same time for everyone. For all other methods (IPTW, censoring and weighting, sequential trials and g-estimation), weights are used to account for censoring.

Let $C_t$ denote a censoring indicator that is set to 1 if the individuals is censored by time $t$ and 0 otherwise. Individuals were censored at time $t$ if they died, received organ transplant, or initiated particular treatments by time $t$, or if their last recorded annual review visit was at time $t$.

The censoring weights used in IPTW, censoring and weighting, and sequential trials ($CENS.w$) were then defined as:

$$CENS.w1_t = \frac{\prod_{j=0}^{t} \Pr(C_{j+1} = c_{j+1} | \bar{c}_j = \bar{0}, L_B = l)}{\prod_{j=0}^{t} \Pr(CENS_{j+1} = c_{j+1} | \bar{c}_j = \bar{0}, \bar{Z}_{j-1} = \bar{z}_{j-1}, L_B = l, \bar{L}_j = \bar{l}_j)} \quad \text{S.17}$$

The censoring weights used in G-estimation were defined as[5]:

$$CENS.w2_t = \frac{I(C_T = 0)}{\prod_{j=t+1}^{T} \Pr(C_j = 0 | \overline{cens}_{j-1} = \bar{0}, \bar{Z}_{j-1} = \bar{z}_{j-1}, L_B = l, \bar{L}_{j-1} = \bar{l}_{j-1})} \quad \text{S.18}$$

Conditional probabilities required for the censoring weights were estimated using logistic regression. All weights were truncated at the 10th and 90th percentiles.

*Inverse-probability-of-treatment weighting*

The stabilised inverse-probability-of-treatment weights at time $t$ were defined as:

$$IPT.w_t = \frac{\prod_{j=0}^{t} \Pr(Z_j = z_j | \bar{Z}_{j-1} = \bar{z}_{j-1}, L_B = l)}{\prod_{j=0}^{t} \Pr(Z_j = z_j | \bar{Z}_{j-1} = \bar{z}_{j-1}, L_B = l, \bar{L}_j = \bar{l}_j)} \quad \text{S.19}$$

The probabilities required for these weights were obtained using logistic regression. The final weights used for each individual was defined as the produce of the inverse-probability-of-treatment weights and the censoring weights.

The following marginal structural model (MSM) was then fitted in the weighted population:

$$Y_{t+1}^{\bar{a}_t} = \beta_0 + \sum_{j=1}^{t} \sum_{c=1}^{3} \beta_{cj} I(z_j = c) + \beta_B L_B + \beta_t t + \varepsilon_{it}, t = 0, \ldots, 4 \quad \text{S.20}$$

*Censoring and weighting and the sequential trials approach*

When using censoring and weighting, individuals were censored at the time they deviated from the treatment strategy they were following at time 0. For each individual at each time point, we estimated the inverse probability of remaining uncensored at that time; these are equivalent to the inverse-probability-of-treatment weights defined in equation S.19. As in the IPTW approach, the final weights used to weight uncensored individuals were a product of the inverse-probability-of-treatment weights, censoring (due to death, transplant, initiating particular treatments, or administrative end of follow-up) weights. The MSM defined in equation S.20 was then fitted in the censored and weighted population to estimate treatment effects.

The sequential trials approach included five "trials" which started at times $t = 0,1,2,3,4$. At time $t$, individuals were eligible for inclusion if they had been taking DNase, but not hypertonic saline, for at least two consecutive years. All trials were pooled into one and the censoring and weighting analysis described above was used to analyse the pooled trials, but with an indicator for "trial" included in the treatment model and the MSM.

*G-formula*

To implement the G-formula, the following model was specified for the outcome:

$$E(Y_{t+1}^{\bar{a}_t}) = \beta_0 + \sum_{j=1}^{t}\sum_{c=1}^{3} \boldsymbol{\beta_{cj}}I(z_j = c) + \boldsymbol{\beta_B L_B} + \boldsymbol{\beta_t L_t} + \beta_t t + \varepsilon_{it} \qquad \text{S.21}$$

Treatment at time $t$ was modelled using multinomial regression with a categorical variable denoting treatment group as the outcome ($Z_t$) and $Z_{t-1}$, $\boldsymbol{L_t}$, $\boldsymbol{L_B}$ and $t$ as predictors. The time variable, $t$ was treated as a continuous predictor.

Individual models were specified for each time-varying covariate in $\boldsymbol{L_t}$ with $Z_{t-1}$, $\boldsymbol{L_{t-1}}$, $\boldsymbol{L_B}$ and $t$ as predictors. Linear regression was used for the continuous covariates, logistic regression was used for binary covariates and multinomial regression was used for categorical covariates.

The gfoRmula[6] package was used to simulate datasets of size 10,000 under each treatment strategy of interest. Relevant end of follow-up outcomes were used to estimate treatment effects.

*G-estimation*

G-estimation was conducted using modified code for the gestMultiple command in the R package gesttools[7]. The code is available on GitHub: https://github.com/danieltompsett/gesttools/blob/master/gestMultiple.R (last accessed: 21/03/2024). We required models for the propensity score, outcome at each time point and for the censoring indicator. Censoring weights were as defined in Equation S.18. Outcome models for each time point were defined as in Equation S.21 and were fit using Generalising Estimating Equations with an independent working correlation matrix in the weighted population. The propensity score was estimated using multinomial regression:

$$\ln\left(\frac{P(Z_t=z_t)}{P(Z_t=0)}\right) = \beta_{0,z} + \sum_{c=1}^{3}\beta_{cz}I(z_{t-1}=c) + \boldsymbol{\beta_{L,}L_t} + \boldsymbol{\beta_B L_B}, \text{ for } z = 1,2,3 \qquad \text{S.22}$$

Where $Z_t$ is defined in Equation (S.15).

*Missing data*

We excluded people who had missing treatment, outcome, or covariate data at baseline. Last one carried forward was then used to impute all other missing data. This approach was chosen because it is a simple way to create one complete dataset which could be used in all analysis methods. This allowed us to focus on comparing the results between different methods for time-varying confounding, without considering what impact missing data had on the performance of the methods. In an analysis where the aim is to estimate the effect of treatment as accurately and as precisely as possible (rather than to compare results between methods), this is not necessarily the best approach to handling missing data. Careful consideration of the reasons for missingness, missing data patterns, and the assumptions made in different missing data methods are needed to inform the best approach to handling missing data[8,9].

## S.2.2 Additional results

Figure S.11 shows how individuals were selected in the study sample and Table S.8 summarises the characteristics at baseline, of the study population, by treatment combination they were observed to be using in the first year of follow-up.

*Figure S.12: Flowchart of participant selection into the study sample*

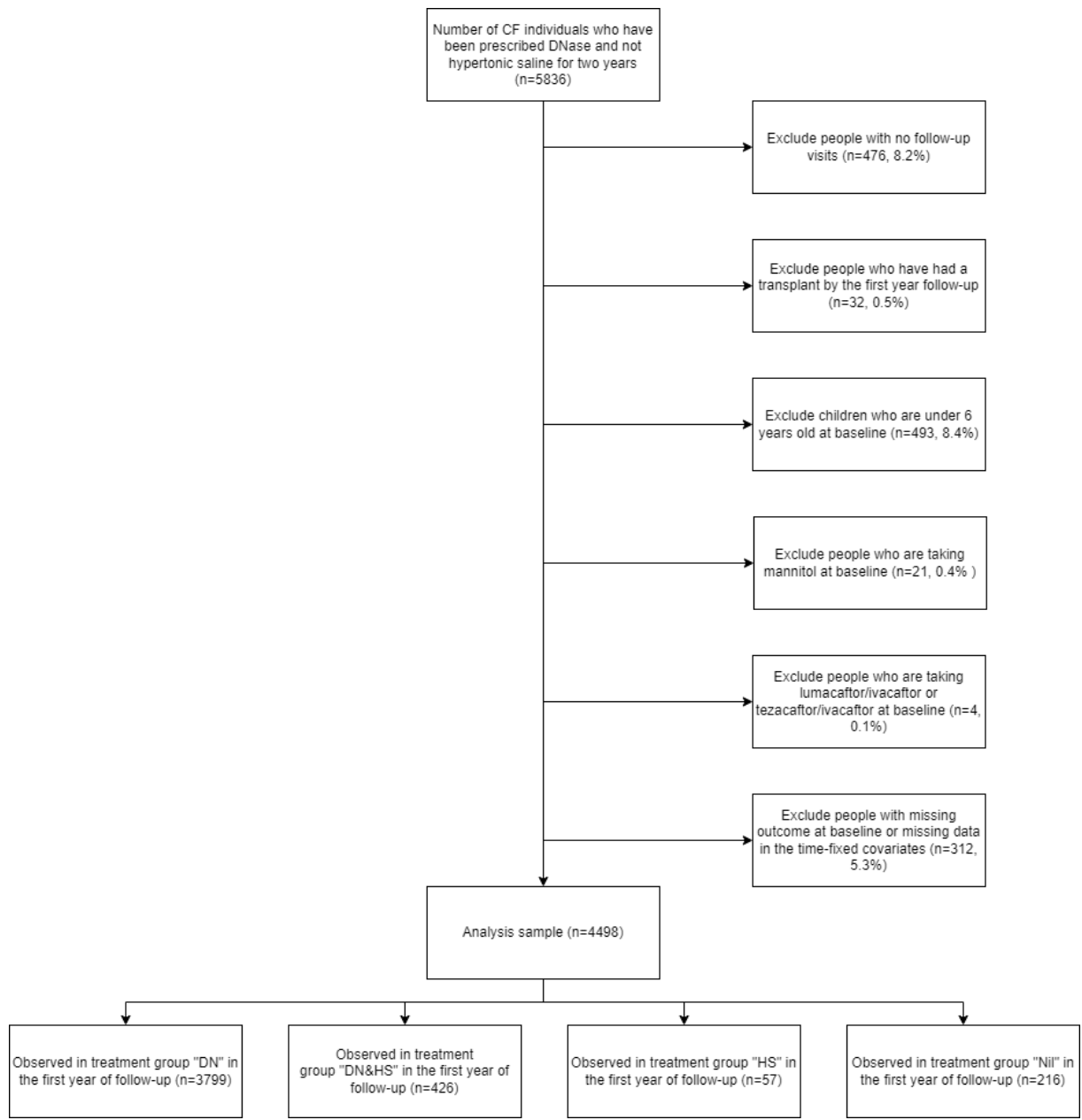

Table S.8: Summary of characteristics at baseline overall and by treatment combination observed in the first year of follow-up. Continuous variables are summarised using mean (standard deviation (SD)) and categorical variables are summarised using numbers (%).

|  | DN (N=3799) | DN&HS (N=426) | HS (N=57) | Nil (N=216) | Whole cohort (N=4498) |
|---|---|---|---|---|---|
| Female, n (%) | 1733 (45.6%) | 223 (52.3%) | 31 (54.4%) | 106 (49.1%) | 2093 (46.5%) |
| Age, Mean (SD) | 21.3 (11.6) | 18.7 (10.3) | 19.7 (9.6) | 24.2 (11.0) | 21.1 (11.5) |
| Genotype risk group*, n(%) |  |  |  |  |  |
| High | 2994 (78.8%) | 358 (84.0%) | 46 (80.7%) | 154 (71.3%) | 3552 (79.0%) |
| Low | 301 (7.9%) | 24 (5.6%) | 3 (5.3%) | 22 (10.2%) | 350 (7.8%) |
| None assigned | 504 (13.3%) | 44 (10.3%) | 8 (14.0%) | 40 (18.5%) | 596 (13.3%) |
| White ethnicity, n (%) | 3653 (96.2%) | 409 (96.0%) | 56 (98.2%) | 208 (96.3%) | 4326 (96.2%) |
| Number of IV days over the past year, n (%) |  |  |  |  |  |
| 0 | 1665 (43.8%) | 151 (35.4%) | 24 (42.1%) | 100 (46.3%) | 1940 (43.1%) |
| 1-14 | 718 (18.9%) | 75 (17.6%) | 12 (21.1%) | 36 (16.7%) | 841 (18.7%) |
| 15-28 | 490 (12.9%) | 66 (15.5%) | 12 (21.1%) | 24 (11.1%) | 592 (13.2%) |
| 29+ | 926 (24.4%) | 134 (31.5%) | 9 (15.8%) | 56 (25.9%) | 1125 (25.0%) |
| IV hospital admissions**, n (%) | 1638 (43.1%) | 220 (51.6%) | 28 (49.1%) | 91 (42.1%) | 1977 (44.0%) |
| FEV1%, Mean (SD) | 70.1 (22.8) | 67.9 (22.7) | 66.2 (17.9) | 66.6 (24.6) | 69.7 (22.8) |
| Rate of decline in FEV1%, Mean (SD) | -1.09 (1.50) | -1.33 (1.63) | -1.48 (1.28) | -1.23 (1.67) | -1.12 (1.52) |
| BMI z-score, Mean (SD) | -0.05 (1.13) | -0.21 (1.13) | -0.31 (0.98) | -0.23 (1.26) | -0.08 (1.13) |
| *P.aeruginosa* infection***, n(%) | 2286 (60.2%) | 258 (60.6%) | 39 (68.4%) | 144 (66.7%) | 2727 (60.6%) |
| *Staphylococcus* infection***, n (%) | 1529 (40.2%) | 169 (39.7%) | 22 (38.6%) | 95 (44.0%) | 1815 (40.4%) |
| Nontuberculous, mycobacteria infection ***, n (%) | 173 (4.6%) | 25 (5.9%) | 4 (7.0%) | 11 (5.1%) | 213 (4.7%) |
| Pancreatic insufficiency, n (%) | 3367 (98.9%) | 392 (92.0%) | 48 (84.2%) | 186 (86.1%) | 3993 (88.8%) |
| Prescribed Ivacaftor, n (%) | 42 (1.1%) | 4 (0.9%) | 0 (0.0%) | 4 (1.9%) | 50 (1.1%) |

Nil: Drop DNase and do not start hypertonic saline; HS: Drop DNase and start hypertonic saline; DN: Continue DNase and do not start hypertonic saline; DN&HS: Continue DNase and start hypertonic saline
*High-risk and low-risk genotype classifications previously defined in [22]. Genotypes which do not fall within either category were labelled "none assigned"
**IV hospital admissions: number of people with at least one IV hospital admission since the last review
***Infection data: indicator for any positive culture since the last review
.